\documentclass{article}

\PassOptionsToPackage{numbers,sort&compress}{natbib}

\usepackage{amsmath, amssymb}
\usepackage{algorithm}
\usepackage{algpseudocode}
\usepackage{graphicx}
\usepackage{subcaption}

\usepackage{booktabs}
\usepackage{multirow} 
\usepackage{multicol}
\usepackage{array}
\usepackage{caption}
\usepackage{caption}
\usepackage{enumitem}
\usepackage{tabularx}

\usepackage[final]{neurips_2024}    

\usepackage[numbers,sort&compress]{natbib}
\usepackage[backref=page, colorlinks=true, linkcolor=blue, citecolor=blue, urlcolor=blue]{hyperref}

\usepackage[colorlinks=true, linkcolor=blue, citecolor=blue, urlcolor=blue]{hyperref}

\usepackage[utf8]{inputenc} 
\usepackage[T1]{fontenc}    
\usepackage{hyperref}       
\usepackage{url}            
\usepackage{booktabs}       
\usepackage{amsfonts}       
\usepackage{nicefrac}     
\usepackage{microtype}      
\usepackage{xcolor}

\title{Hybrid Learning and Optimization methods for solving Capacitated Vehicle Routing Problem}

\author{
  Monit Sharma \\
  School of Computing and Information Systems \\
  Singapore Management University \\
  \texttt{monitsharma@smu.edu.sg} \\
  \And
  Hoong Chuin Lau  \\
  School of Computing and Information Systems \\
  Singapore Management University \\
  \texttt{hclau@smu.edu.sg} \\
}

\begin{document}

\maketitle

\begin{abstract}
   The Capacitated Vehicle Routing Problem (CVRP) is a fundamental NP-hard problem in logistics. Augmented Lagrangian Methods (ALM) for solving CVRP performance depends heavily on well-tuned penalty parameters. In this paper, we propose a hybrid optimization approach that integrates deep reinforcement learning (RL) to automate the selection of penalty parameter values within both classical (RL-C-ALM) and quantum-enhanced (RL-Q-ALM) ALM solvers.

Using Soft Actor-Critic, our approach learns penalty values from CVRP instance features and constraint violations. In RL-Q-ALM, subproblems are encoded as QUBOs and solved using Variational Quantum Eigensolvers (VQE). The agent learns across episodes by maximizing solution feasibility and minimizing cost.

Experiments show that RL-C-ALM outperforms manually tuned ALM on synthetic and benchmark CVRP instances, achieving better solutions with fewer iterations. Also, RL-Q-ALM matches classical solution quality on small instances but incurs higher runtimes due to quantum overhead. Our results highlight the potential of combining RL with classical and quantum solvers for scalable, adaptive combinatorial optimization \footnote{Code available at: \url{https://github.com/SMU-Quantum/adaptive_quantum_cvrp}}.

\end{abstract}

\section{Introduction}

The Capacitated Vehicle Routing Problem (CVRP) is a canonical problem in combinatorial optimization, with extensive applications across logistics, transportation, and supply chain design \cite{toth2002vehicle}. Given a set of geographically distributed customers and a fleet of identical-capacity vehicles, the objective is to determine a set of vehicle routes, each starting and ending at a central depot, such that each customer is visited exactly once, vehicle capacities are not exceeded, and the total cost, typically the aggregate travel distance, is minimized \cite{dantzig1959truck,clarke1964scheduling}.

Despite decades of algorithmic development, CVRP remains computationally challenging due to its combinatorial structure and NP-hardness \cite{lenstra1981complexity}. Exact approaches, such as branch-and-cut, quickly become impractical as the number of customers grows \cite{lysgaard2004new,naddef2002branch}, while heuristic and metaheuristic methods provide tractable approximations at the expense of optimality guarantees \cite{gendreau1994tabu, laporte2002classical}. Recent years have seen the emergence of data-driven methods, including reinforcement learning (RL), for solving routing problems by learning instance-aware heuristics \cite{nazari2018reinforcement,wu2021learning}. In parallel, quantum optimization, via variational quantum algorithms and quantum annealing, has been proposed as a future-facing paradigm for solving hard discrete problems \cite{fitzek2024applying}, albeit currently limited by hardware constraints \cite{sharma2025comparative}.

In this work, we explore the integration of RL and quantum optimization within a classical optimization framework based on the Augmented Lagrangian Method (ALM) \cite{Hestenes1969,Powell1969}. We propose and systematically evaluate four solvers for CVRP:
\begin{itemize}
    \item \textbf{C-ALM:} a classical ALM approach adapted for CVRP, handling routing and capacity constraints via Lagrangian multipliers and penalty terms;
    \item \textbf{RL-C-ALM:} an RL-guided ALM where a learned policy selects penalty parameters to accelerate convergence and improve solution quality;
    \item \textbf{Q-ALM:} a quantum-enhanced ALM wherein subproblems are mapped to QUBOs and solved via the Variational Quantum Eigensolver (VQE) \cite{sharma2025cutting};
    \item \textbf{RL-Q-ALM:} a novel fusion of RL and quantum solvers, where RL selects parameters for a quantum-powered ALM loop.
\end{itemize}

This study benchmarks these paradigms on synthetic and standard CVRP datasets \cite{Uchoa2017}, assessing solution quality, feasibility, convergence, and scalability. Our contributions provide insights into the comparative utility of RL and quantum components, and establish a robust framework for hybrid optimization under realistic constraints.

\subsection{The Capacitated Vehicle Routing Problem}

Let \( V = \{0, 1, \dots, n\} \) denote the set of nodes, where \( 0 \) is the depot and \( \{1, \dots, n\} \) are customers. Each customer \( i \in \{1, \dots, n\} \) has a nonnegative demand \( d_i \), and each of the \( |K| \) vehicles has identical capacity \( Q \). The travel cost between nodes \( i \) and \( j \) is denoted \( c_{ij} \).

A feasible CVRP solution comprises a set of routes such that:
\begin{enumerate}
    \item Each customer is visited exactly once by a single vehicle;
    \item Each vehicle starts and ends at the depot;
    \item The total demand on each route does not exceed \( Q \).
\end{enumerate}
The objective is to minimize the total cost of all vehicle routes.

We present a three-index vehicle-flow ILP formulation with Miller-Tucker-Zemlin (MTZ) subtour elimination constraints, following \cite{toth2002vehicle}.

\paragraph{Decision variables.}
\[
x_{ijk} =
\begin{cases}
1 & \text{if vehicle } k \text{ travels from node } i \text{ to node } j, \\
0 & \text{otherwise}
\end{cases}
\quad \forall i,j \in V, i \ne j, \forall k \in K
\]
\[
u_{ik} = \text{cumulative load on vehicle } k \text{ just after visiting customer } i, \quad \forall i \in \{1,\dots,n\}, \forall k \in K
\]

\paragraph{Objective.}
\[
\min \sum_{k \in K} \sum_{i \in V} \sum_{\substack{j \in V \\ j \ne i}} c_{ij} x_{ijk}
\]

\paragraph{Constraints.}
\begin{align}
\text{(Customer visit)} &\quad \sum_{k \in K} \sum_{\substack{i \in V \\ i \ne j}} x_{ijk} = 1, && \forall j \in \{1, \dots, n\} \label{eq:visit} \\
\text{(Depot depart)} &\quad \sum_{j = 1}^n x_{0jk} = 1, && \forall k \in K \label{eq:depart} \\
\text{(Depot return)} &\quad \sum_{i = 1}^n x_{i0k} = 1, && \forall k \in K \label{eq:return} \\
\text{(Flow conservation)} &\quad \sum_{\substack{i \in V \\ i \ne j}} x_{ijk} = \sum_{\substack{l \in V \\ l \ne j}} x_{jlk}, && \forall j \in \{1, \dots, n\}, \forall k \in K \label{eq:flow} \\
\text{(Capacity)} &\quad \sum_{i = 1}^n d_i \sum_{\substack{j \in V \\ j \ne i}} x_{ijk} \le Q, && \forall k \in K \label{eq:capacity} \\
\text{(MTZ subtour elimination)} &\quad u_{ik} + d_j \le u_{jk} + Q(1 - x_{ijk}), && \forall i,j \in \{1,\dots,n\}, i \ne j, \forall k \in K \label{eq:mtz1} \\
&\quad d_i \le u_{ik} \le Q, && \forall i \in \{1,\dots,n\}, \forall k \in K \label{eq:mtz2} \\
&\quad u_{ik} \ge d_i \sum_{j=0}^n x_{jik}, && \forall i \in \{1,\dots,n\}, \forall k \in K \label{eq:mtz3}
\end{align}

Constraint \eqref{eq:mtz1} enforces that if vehicle \( k \) travels from \( i \) to \( j \), then the load at \( j \) must be at least the load at \( i \) plus \( j \)'s demand. Constraint \eqref{eq:mtz2} bounds the load variables, and \eqref{eq:mtz3} ensures that \( u_{ik} \) is activated only if customer \( i \) is visited by vehicle \( k \). Together, these eliminate disconnected subtours as well as enforce vehicle capacity along routes. Strictly speaking therefore, constraint \eqref{eq:capacity} is redundant; however it can still be useful to tighten the LP relaxation. 

\paragraph{Domain.}
\[
x_{ijk} \in \{0,1\}, \quad \forall i,j \in V, i \ne j, \forall k \in K; \quad u_{ik} \in \mathbb{R}_+, \quad \forall i \in \{1,\dots,n\}, \forall k \in K
\]

\paragraph{Complexity.}
The CVRP is strongly NP-hard \cite{lenstra1981complexity}. Its search space grows factorially with the number of customers and combinatorially with the number of vehicles. As a result, exact algorithms become computationally intractable for instances beyond moderate sizes (e.g., \( n \ge 50 \)), motivating the need for approximation, learning-based, and hybrid strategies.

While this paper focuses on the canonical single-depot CVRP for clarity and benchmarking, the problem is part of a broader family of VRPs, including Time Windows (VRPTW) \cite{vidal2020concise}, Multiple Depots (MDVRP) \cite{gulczynski2011multi}, and Split Deliveries (SDVRP) \cite{feillet2006vehicle}. Advances in solving the base CVRP often transfer to these richer variants, providing foundational tools for real-world logistics optimization.

\subsection{Hybrid Solvers for the Capacitated Vehicle Routing Problem}

This paper investigates four distinct paradigms for solving the Capacitated Vehicle Routing Problem (CVRP), emphasizing a rigorous comparative evaluation under a unified experimental framework. The approaches differ in their computational foundations, ranging from classical optimization to quantum computing, and offer insights into the capabilities and limitations of emerging hybrid solvers.

\begin{enumerate}
    \item \textbf{Classical Augmented Lagrangian Method (C-ALM):} A deterministic, penalty-based optimization approach that formulates CVRP as a constrained problem and employs augmented Lagrangian techniques to enforce routing and capacity constraints.
    
    \item \textbf{Quantum-Augmented Lagrangian Method (Q-ALM):} A hybrid algorithm that retains the classical ALM framework but replaces the subproblem solver with quantum optimization. Specifically, each routing subproblem is reformulated as a Quadratic Unconstrained Binary Optimization (QUBO) instance and solved using quantum variational algorithms such as the Variational Quantum Eigensolver (VQE) \cite{peruzzo2014variational} or the Quantum Approximate Optimization Algorithm (QAOA) \cite{farhi2014quantum}.
    
    \item \textbf{Reinforcement Learning–Tuned Classical ALM (RL-C-ALM):} A learning-augmented solver in which a deep RL agent learns to adjust penalty parameters and Lagrange multipliers for ALM. This tuning accelerates convergence and improves feasibility in classical settings.
    
    \item \textbf{Reinforcement Learning–Tuned Quantum ALM (RL-Q-ALM):} A novel quantum-classical hybrid that combines RL-guided parameter tuning with quantum optimization. The RL agent predicts penalty weights, steering the quantum solver toward feasible solutions in the presence of noisy or approximate circuits.
\end{enumerate}

The novelty lies not merely in the design of each individual solver but in the systematic and fair comparison across all four paradigms on standard CVRP instances. The introduction of RL-guided quantum solvers for CVRP constitutes a forward-looking contribution, positioned at the intersection of quantum optimization and machine learning.

\subsection{Key Contributions}

This work resides at the intersection of Operations Research, quantum computing, and reinforcement learning. It addresses a key question: How can emerging quantum and learning-based techniques be meaningfully integrated into classical algorithmic frameworks to solve structured, NP-hard problems such as CVRP?

While most existing quantum optimization studies focus on synthetic or unstructured QUBOs, our approach grounds quantum methods in a real-world logistics setting, making the quantum contribution more applicable and measurable. Furthermore, by embedding quantum solvers into ALM and tuning them with RL, we introduce a flexible architecture that accommodates quantum noise, adapts to varying problem sizes, and maintains constraint feasibility, factors critical for near-term quantum applicability.

Our contributions are as follows:

\begin{itemize}
    \item \textbf{Methodological Innovation:}
    \begin{itemize}
        \item \textbf{C-ALM:} A tailored formulation of the Augmented Lagrangian Method for CVRP, leveraging structure-aware penalty terms and multiplier updates to handle routing, capacity, and feasibility constraints.
        \item \textbf{RL-C-ALM:} Design of a deep RL agent (Soft Actor-Critic) that learns optimal penalty settings for ALM based on instance features and violation patterns, improving convergence and solution quality over fixed-parameter baselines.
        \item \textbf{Q-ALM:} An integration of quantum optimization into the ALM framework. CVRP subproblems are encoded as QUBOs and solved via quantum variational algorithms (e.g., VQE), providing a hybrid pipeline that explores the feasibility of quantum-assisted constraint satisfaction.
        \item \textbf{RL-Q-ALM:} The first integration of reinforcement learning and quantum variational solvers for CVRP. Penalty parameters are predicted offline by an RL policy and injected into QUBO formulations solved by VQE or QAOA, enabling scalable quantum-assisted constraint satisfaction.
    \end{itemize}
    
    \item \textbf{Empirical Evaluation:} A comprehensive experimental comparison of all four methods across standard CVRP benchmarks, evaluating:
    \begin{itemize}
        \item Solution quality and feasibility
        \item Quantum circuit depth and shot robustness
        \item Generalization across different instance sizes
    \end{itemize}

    \item \textbf{Synergistic Framework Design:} The proposed architecture showcases how classical constraint-handling (ALM), quantum solvers (VQE/QAOA), and RL-based penalty adaptation can be integrated synergistically. In particular, the ALM formulation enables a compact and scalable QUBO encoding of CVRP constraints, reducing the qubit footprint and improving tractability for noisy intermediate-scale quantum (NISQ) devices.

    \item \textbf{Cross-Domain Transferability:} The RL-Quantum framework developed here is not limited to CVRP. It generalizes to other combinatorial problems with hard constraints (e.g., scheduling, portfolio optimization), offering a blueprint for extending RL-tuned QUBO solvers to a broader range of real-world tasks.
\end{itemize}

In summary, we demonstrate that interdisciplinary methods, when properly aligned, can provide practical advances even under the limitations of near-term quantum hardware. Our results underscore the importance of joint design across solver architecture, constraint formulation, and parameter adaptation.

\section{Literature Review}

\subsection{Augmented Lagrangian Methods for CVRP}

Augmented Lagrangian Methods (ALM) provide a powerful mathematical framework for constrained optimization, combining the strength of Lagrange duality with penalty-based regularization ~\cite{Hestenes1969,Powell1969,HeldKarp1971,fisher1981lagrangian,birgin2014practical}. Their adaptation to combinatorial settings such as the VRP requires careful reformulation and decomposition strategies. This section reviews foundational results on ALM, its applications to discrete and routing problems, and the architectural choices made when tailoring it to CVRP.

\subsubsection{ALM in Constrained and Combinatorial Optimization}

ALM reformulates a constrained problem into a sequence of unconstrained (or partially constrained) subproblems by augmenting the Lagrangian with quadratic penalties for constraint violations. The method alternates between minimizing this augmented Lagrangian and updating multipliers to reflect constraint duality. Compared to pure penalty methods, ALM avoids the numerical instabilities caused by excessively large penalties by leveraging multiplier feedback.

The classical ALM formulation for minimizing a function \( f(x) \) subject to equality constraints \( g_i(x) = 0 \) takes the form:
\[
\mathcal{L}_\rho(x, \lambda) = f(x) + \sum_{i} \lambda_i g_i(x) + \frac{\rho}{2} \sum_i g_i(x)^2,
\]
with multipliers updated via:
\[
\lambda_i^{k+1} = \lambda_i^k + \rho_k g_i(x^k).
\]

For convex continuous problems, convergence theory is well-developed. Under regularity conditions such as constraint qualification and strict complementarity, ALM achieves linear or even superlinear convergence \cite{Hestenes1969, Powell1969}.

The extension to combinatorial problems, including discrete or integer optimization, is more recent and involves decomposition techniques or dual relaxations \cite{liao2025bundlebasedaugmentedlagrangianframework}. In routing problems like CVRP, ALM can be used to relax global constraints (e.g., assignment, capacity), allowing decomposition into tractable subproblems. Notably, this aligns with strategies in column generation and Lagrangian decomposition, which solve pricing subproblems (e.g., ESPPRC) in VRP variants \cite{Bai2020, SONG2024121511}.

Recent work by Song et al.~\cite{SONG2022108736} applies ALM to nonlinear VRP formulations using an Alternating Direction Method of Multipliers (ADMM)-style decomposition. This synergy between ALM and column generation provides a foundation for scalable hybrid solvers.

\subsubsection{Architectural Design: Tailoring ALM to CVRP}

A successful application of ALM to CVRP depends on effective decomposition, tractable subproblem formulations, and robust penalty and multiplier update rules.

\paragraph{Decomposition Strategies.}

Let us denote the primary CVRP constraints:
\begin{itemize}
    \item Visit constraint: each customer must be visited once.
    \item Capacity constraint: total demand per route must not exceed capacity.
    \item Routing feasibility: routes must be connected tours.
\end{itemize}

Several decomposition strategies can be considered:

\begin{itemize}
    \item \textbf{Relaxing Assignment Constraints.} If customer-to-vehicle assignments are dualized, the problem decomposes across vehicles. Each subproblem becomes a capacity-constrained routing task, typically an instance of the Elementary Shortest Path Problem with Resource Constraints (ESPPRC). This decomposition is aligned with classical VRP dual decomposition strategies \cite{NIU201870}.
    
    \item \textbf{Relaxing Routing/Sequencing Constraints.} Alternatively, relaxing subtour elimination or ordering constraints leads to subproblems resembling generalized assignment problems or capacitated set cover problems, with penalties enforcing feasible tours.
    
    \item \textbf{ADMM-Style Splitting.} If the problem is separable in variable blocks, e.g., binary route indicators and continuous load variables, an ADMM variant of ALM can be applied. This enables alternating updates between route selection and load balancing under relaxed coupling constraints.
\end{itemize}

\paragraph{Subproblem Formulation.}

The complexity of the resulting subproblems depends on the relaxation. ESPPRC subproblems remain NP-hard, but can be solved using label-setting or dynamic programming for moderate instances \cite{feillet2004exact, lozano2016exact}. There is a trade-off between subproblem tractability and dual bound strength: simpler subproblems accelerate iterations but may slow ALM convergence, whereas tighter subproblems are costlier but improve dual progress \cite{righini2006symmetry}.

\subsubsection{Advanced Multiplier and Penalty Update Strategies}

Multiplier updates are typically driven by constraint violation:
\[
\lambda_i^{k+1} = \lambda_i^k + \rho_k \, g_i(x^k),
\]
where \( \lambda_i \) corresponds to a CVRP constraint (e.g., capacity overload or unvisited customers) and \( \rho_k \) is the penalty parameter \cite{Hestenes1969, Powell1969, wright1999numerical}.

Managing \( \rho_k \) is crucial. If too low, constraint satisfaction is slow; if too high, subproblems become numerically unstable. Adaptive strategies are recommended:
\begin{itemize}
    \item Increase \( \rho_k \) when constraint violation stagnates.
    \item Decrease \( \rho_k \) if subproblem solvers fail to converge.
    \item Employ heuristic or learning-based updates to modulate \( \rho_k \) dynamically \cite{wright1999numerical, birgin2014practical}.
\end{itemize}

These updates are typically guided by dual residuals or feasibility violation norms \cite{boyd2011distributed}. In our later sections, we show how reinforcement learning can further enhance this adaptive tuning process by learning optimal penalty schedules across CVRP instances \cite{stooke2020responsive}.

\subsubsection{ALM as a Warm-Start Strategy}

Even when ALM does not reach full convergence, the intermediate feasible solutions often exhibit high quality and low constraint violation. In vehicle routing, Lagrangian-relaxation has a long history of producing strong seeds that feed construction/improvement heuristics and exact methods \cite{fisher1981generalized, kallehauge2006lagrangian,toth2014vehicle}. 

Concretely, intermediate ALM solutions can initialize:
\begin{itemize}
    \item Local improvement (e.g., 2-opt) to refine tours quickly. \cite{croes1958method}.
    \item Savings-based constructive heuristics (Clarke–Wright) to rebuild routes around a good assignment \cite{pichpibul2013heuristic, paessens1988savings,lysgaard1997clarke}.
    \item Metaheuristics (GA, SA, LNS) or branch-and-price/column-generation frameworks that benefit from quality starts and good dual information \cite{toth2014vehicle,kallehauge2006lagrangian}.
\end{itemize}

Thus, ALM serves not only as a solver but also as a unifying scaffold for hybrid pipelines (learning + classical), where its iterates seed heuristics and provide guidance signals (e.g., penalties, multiplier trajectories) that can be used as training supervision or reward shaping for RL agents \cite{toth2014vehicle}.

\subsection{Reinforcement Learning for Adaptive Parameter Tuning in Optimization}

The effectiveness of optimization frameworks such as the Augmented Lagrangian Method (ALM) in solving structured problems like the Capacitated Vehicle Routing Problem (CVRP) hinges significantly on the proper calibration of internal hyperparameters, most notably, penalty coefficients. Traditional practices rely on heuristic or static schedules to set these values \cite{bollapragada2023adaptive, xiao2022class}. However, such approaches are typically instance-dependent, manually intensive, and sensitive to initialization. Reinforcement Learning (RL) has recently emerged as a promising paradigm for automating this tuning process, enabling dynamic, data-driven adjustment of optimization parameters to improve robustness and convergence across diverse problem instances \cite{zeng2022reinforcement, li2021augmented,chen2025reinforcementlearningconstrainedbeam}.

\subsubsection{RL for Hyperparameter and Penalty Coefficient Optimization}

The task of determining optimal penalty parameters in constrained optimization algorithms can be formulated as a sequential decision-making problem, naturally fitting the reinforcement learning framework. In this setting, an RL agent interacts with the optimization algorithm as an environment. The agent observes a state \( s_t \) comprising instance features (e.g., number of customers, degree of infeasibility, multiplier gradients), and selects an action \( a_t \) corresponding to penalty parameter values (e.g., \( \rho_j, \sigma_k \)). The reward \( r_t \) is based on downstream optimization performance, such as:
\begin{itemize}
    \item Objective cost after a fixed number of ALM iterations,
    \item Degree of constraint violation,
    \item Speed or stability of convergence.
\end{itemize}

This enables the agent to learn an adaptive policy \( \pi(a \mid s) \) that generalizes across problem instances, removing the need for instance-specific manual tuning \cite{ji2023updatemodelconstrainedmodelbased}.

In constrained optimization problems (COPs), the conventional penalty-based formulation:
\[
r_{\text{total}} = r_{\text{objective}} + p_{\text{constraint}},
\]
requires careful balancing. If penalties are set too low, constraint violations persist; if too high, the optimization may over-prioritize feasibility at the cost of solution quality or numerical stability \cite{Farooq_2024}. Reinforcement learning provides a principled framework to automate this trade-off, allowing penalty parameters to evolve dynamically as the optimization progresses.

\subsubsection{RL in the Context of Constrained Optimization Algorithms}

Beyond scalar penalty adjustment, reinforcement learning has been integrated into optimization architectures more broadly. Lagrangian-based RL methods formulate constrained optimization as a saddle-point problem and learn to update Lagrange multipliers directly, e.g., Predictive Lagrangian Optimization (PLO) \cite{zhang2025predictive} applies a model-predictive control (MPC) lens to anticipate constraint violations and adjust dual variables proactively.

Our approach adopts a hybrid design philosophy. We retain the classical ALM update rule for Lagrange multipliers but embed a learned RL policy to initialize and guide the selection of penalty parameters. This allows the system to benefit from the theoretical structure of ALM while gaining adaptivity from the RL agent. Compared to inference-time constraint masking strategies (e.g., RL-Constrained Beam Search \cite{chen2025reinforcementlearningconstrainedbeam}), our method enforces constraints through optimization and enables fine-grained penalty adaptation throughout.

\section{The Classical Augmented Lagrangian Method for CVRP (C-ALM)}
\label{sec:calm}

This section details the foundational classical optimization framework, the Classical Augmented Lagrangian Method (C-ALM), which serves as the baseline for the learning-based and quantum-enhanced methods developed in this work. The C-ALM approach formulates the Capacitated Vehicle Routing Problem (CVRP) as a constrained optimization problem and utilizes augmented Lagrangian techniques to handle routing and capacity constraints. By establishing a clear understanding of this core algorithmic structure, a solid foundation is laid for appreciating the subsequent enhancements introduced by reinforcement learning and quantum computing.

\subsection{Augmented Lagrangian Formulation for CVRP}

We apply Augmented Lagrangian Method (ALM) where the following CVRP constraints are relaxed:
\begin{enumerate}
    \item \textbf{Customer Visit Constraints:} Each customer must be visited exactly once.
    \item \textbf{Vehicle Capacity Constraints:} Each route must not exceed vehicle capacity.
\end{enumerate}
These relaxed constraints are incorporated into the objective via linear (dual) and quadratic (penalty) terms. The augmented objective is then minimized over feasible routes.

Let \( S \) denote a candidate solution composed of vehicle routes \( \{R_1, \dots, R_m\} \), where \( m \le K_{\max} \). The base cost is:
\[
C(S) = \sum_{k=1}^m \text{cost}(R_k).
\]

\subsubsection{Customer Visit Constraints}

Let \( v_j(S) \) denote the number of times customer \( j \in C \) is visited. The relaxed constraint contributes:
\[
\lambda_j (v_j(S) - 1) + \frac{1}{2} \rho_j (v_j(S) - 1)^2,
\]
where \( \lambda_j \) and \( \rho_j \) are the Lagrange multiplier and penalty parameter for customer \( j \).

\subsubsection{Vehicle Capacity Constraints}

Let \( L(R_k) = \sum_{i \in R_k,\, i \ne 0} d_i \) denote the total load on route \( R_k \), and define:
\[
h_k(S) = \max\left(0, L(R_k) - Q\right).
\]
The relaxed constraint adds:
\[
\mu_k h_k(S) + \frac{1}{2} \sigma_k h_k(S)^2,
\]
where \( \mu_k \) and \( \sigma_k \) are the Lagrange multiplier and penalty for the capacity constraint on route \( R_k \).

\subsubsection{Augmented Objective Function}

The full augmented Lagrangian objective is:
\begin{equation}
L_{\text{aug}}(S) = C(S) + \sum_{j \in C} \left[ \lambda_j (v_j(S) - 1) + \frac{1}{2} \rho_j (v_j(S) - 1)^2 \right]
+ \sum_{k=1}^{m} \left[ \mu_k h_k(S) + \frac{1}{2} \sigma_k h_k(S)^2 \right].
\end{equation}
This function is iteratively minimized within the ALM loop, while multipliers and penalty parameters are updated based on constraint violations.

\subsection{Subproblem Generation: Route Construction via ESP}

At each ALM iteration, the optimizer generates up to \( K_{\max} \) vehicle routes that approximately minimize \( L_{\text{aug}} \). Since constraints are relaxed, the subproblem of generating a route reduces to an instance of the Elementary Shortest Path (ESP) problem with modified arc costs.

To construct a route \( P = (0, p_1, \dots, p_s, 0) \), we define the cost:
\begin{equation}
\text{Cost}(P) = \sum_{(i,j) \in P} c_{ij} + \sum_{j \in P \setminus \{0\}} \lambda_j.
\end{equation}
Equivalently, the node-specific reward passed to the ESP solver is:
\[
\text{modified\_reward}(j) = -\lambda_j.
\]
This encourages the ESP solver to include customers with high \( \lambda_j > 0 \) (i.e., under-served), balancing travel cost and constraint satisfaction. To promote diversity, a customer once selected in a route is removed from subsequent subproblem inputs during the same ALM iteration.

Capacity multipliers \( \mu_k \) and penalties \( \sigma_k \) do not affect subproblem routing. They are evaluated globally in the augmented objective after route generation.

The C-ALM subproblem solver avoids subtours by construction. It uses a shortest-path heuristic that generates elementary depot-anchored routes during each iteration. These paths are directly constrained to begin and end at the depot, ensuring connected, non-cyclic solutions without requiring additional penalization. This structural advantage contributes to the higher feasibility and solution quality of classical ALM methods on large CVRP instances, as shown in our experimental results.

\subsection{Multiplier and Penalty Parameter Updates}

After constructing solution \( S^{(t)} \) at iteration \( t \), we update the dual variables and penalty terms.

\subsubsection{Customer Visit Constraints}

Let:
\[
g_j^{(t)} = v_j(S^{(t)}) - 1.
\]
Then, the multiplier update is given as follows:
\begin{equation}
\lambda_j^{(t+1)} = \lambda_j^{(t)} + \rho_j^{(t)} \, g_j^{(t)}.
\label{eq:lambda-update}
\end{equation}
The penalty parameter is increased only if the violation is significant:
\[
\rho_j^{(t+1)} =
\begin{cases}
\min\left( \rho_j^{(t)} \times f_\rho,\, \rho_{\max} \right), & \text{if } |g_j^{(t)}| > \varepsilon_{\text{visit}} \cdot \text{factor}, \\
\rho_j^{(t)}, & \text{otherwise}.
\end{cases}
\label{eq:rho-update}
\]

\subsubsection{Capacity Constraints}

Let:
\[
h_k^{(t)} = \max\left(0, L(R_k^{(t)}) - Q \right).
\]
Then, the multiplier update is given as:
\begin{equation}
\mu_k^{(t+1)} = \max\left(0, \mu_k^{(t)} + \sigma_k^{(t)} \, h_k^{(t)} \right),
\label{eq:mu-update}
\end{equation}
and
\begin{equation}
\sigma_k^{(t+1)} =
\begin{cases}
\min\left( \sigma_k^{(t)} \times f_\sigma,\, \sigma_{\max} \right), & \text{if } h_k^{(t)} > \varepsilon_{\text{cap}} \cdot \text{factor}, \\
\sigma_k^{(t)}, & \text{otherwise}.
\end{cases}
\label{eq:sigma-update}
\end{equation}

\subsubsection*{Hyperparameters}

\begin{itemize}
    \item \( f_\rho, f_\sigma \): multiplicative growth factors for penalties.
    \item \( \rho_{\max}, \sigma_{\max} \): upper bounds on penalty values.
    \item \( \varepsilon_{\text{visit}}, \varepsilon_{\text{cap}} \): convergence thresholds for constraint violations.
    \item Updates apply only if corresponding customers/routes were included in solution \( S^{(t)} \).
\end{itemize}

This iterative scheme ensures that constraint satisfaction improves over time while avoiding excessive penalty inflation that may destabilize subproblem solvers.

\subsection{The C-ALM Algorithm and Convergence}

The concepts described in the preceding subsections are consolidated into the C-ALM procedure, which is formally outlined in Algorithm~\ref{alg:alm-cvrp}. This algorithm provides a complete, step-by-step description of the baseline classical solver, from initialization through iterative route generation and parameter updates to final convergence.

\begin{algorithm}[ht]
\caption{Augmented Lagrangian Method (ALM) for CVRP}
\label{alg:alm-cvrp}
\begin{algorithmic}[1]
\State \textbf{Initialize:}
\ForAll{$j \in C$}
    \State $\lambda_j^{(0)} \gets 0$, \quad $\rho_j^{(0)} \gets \rho_{\text{init}}$
\EndFor
\For{$k = 1, \dots, K_{\max}$}
    \State $\mu_k^{(0)} \gets 0$, \quad $\sigma_k^{(0)} \gets \sigma_{\text{init}}$
\EndFor
\State $t \gets 0$, \quad \texttt{best\_feasible\_solution} $\gets \texttt{None}$

\While{$t < T_{\max}$ \textbf{and} not converged}
    \State $t \gets t + 1$
    \State \textbf{// Route Generation via Subproblem Solving}
    \State \texttt{current\_routes} $\gets \emptyset$; \quad \texttt{covered\_customers} $\gets \emptyset$
    \ForAll{$j \in C$}
        \State \texttt{reward}[j] $\gets -\lambda_j^{(t-1)}$
    \EndFor

    \For{$k = 1, \dots, K_{\max}$}
        \State Solve ESP using \texttt{solve\_esp\_with\_dominance} with:
        \Statex \hspace{\algorithmicindent} rewards = \texttt{reward}, tabu set = \texttt{covered\_customers}
        \If{valid route $R_{\text{new}}$ is found}
            \State Add $R_{\text{new}}$ to \texttt{current\_routes}
            \State Compute $L(R_{\text{new}})$
            \State Add customers in $R_{\text{new}} \setminus \{0\}$ to \texttt{covered\_customers}
            \If{all customers covered}
                \State \textbf{break}
            \EndIf
        \Else
            \State \textbf{break}
        \EndIf
    \EndFor

    \State $S^{(t)} \gets$ solution formed from \texttt{current\_routes}

    \State \textbf{// Evaluation}
    \State Compute cost $C(S^{(t)})$
    \State Check feasibility using \texttt{check\_solution\_feasibility}
    \If{$S^{(t)}$ is feasible \textbf{and} $C(S^{(t)})$ improves best}
        \State \texttt{best\_feasible\_solution} $\gets S^{(t)}$
    \EndIf

    \State \textbf{// Compute Constraint Violations}
    \ForAll{$j \in C$}
        \State $g_j^{(t)} \gets v_j(S^{(t)}) - 1$
    \EndFor
    \ForAll{routes $R_k^{(t)}$}
        \State $h_k^{(t)} \gets \max(0, L(R_k^{(t)}) - Q)$
    \EndFor

    \State \textbf{// Update Multipliers and Penalties}
    \ForAll{$j \in C$}
        \State Update $\lambda_j^{(t)}$, $\rho_j^{(t)}$ via Eqs.~\eqref{eq:lambda-update}--\eqref{eq:rho-update}
    \EndFor
    \ForAll{routes $R_k^{(t)}$}
        \State Update $\mu_k^{(t)}$, $\sigma_k^{(t)}$ via Eqs.~\eqref{eq:mu-update}--\eqref{eq:sigma-update}
    \EndFor

    \State \textbf{// Convergence Check}
    \If{$\max_j |g_j^{(t)}| \leq \varepsilon_{\text{visit}}$ \textbf{and}
         $\max_k h_k^{(t)} \leq \varepsilon_{\text{cap}}$ \textbf{and}
         $S^{(t)}$ is feasible}
        \State \textbf{break}
    \EndIf
\EndWhile

\State \Return \texttt{best\_feasible\_solution}
\end{algorithmic}
\end{algorithm}

The Augmented Lagrangian Method (ALM) terminates once one of the following criteria is satisfied:

\begin{enumerate}
    \item \textbf{Maximum Iterations Reached:} \\
    The algorithm halts if the predefined maximum number of iterations is exceeded:
    \[
    t \geq T_{\max},
    \]
    where \( T_{\max} \) is a user-defined hyperparameter.

    \item \textbf{Feasibility Criteria Satisfied:} \\
    The current solution \( S^{(t)} \) is considered \emph{feasible} if both of the following conditions hold:
    
    \begin{itemize}
        \item \textbf{Customer Visit Constraint:} \\
        Each customer is visited approximately once:
        \[
        \max_{j \in C} \left| v_j(S^{(t)}) - 1 \right| \leq \varepsilon_{\text{visit}},
        \]
        where \( v_j(S^{(t)}) \) is the number of visits to customer \( j \), and \( \varepsilon_{\text{visit}}  \)(convergence tolerance) is the tolerance for visit feasibility.
        
        \item \textbf{Capacity Constraint:} \\
        Each route satisfies the capacity constraint up to a threshold:
        \[
        \max_{k = 1,\dots,m^{(t)}} \left[ \max\left(0, L(R_k^{(t)}) - Q \right) \right] \leq \varepsilon_{\text{cap}},
        \]
        where \( L(R_k^{(t)}) \) is the total demand on route \( R_k^{(t)} \), and \( \varepsilon_{\text{cap}} \)(capacity convergence tolerance) is the allowable capacity violation.
    \end{itemize}
\end{enumerate}

This dual convergence criterion ensures that ALM either halts when a feasible solution is found or continues iterating up to a computational budget. The use of tolerance-based stopping criteria allows for graceful termination in near-feasible regimes, critical in settings where exact feasibility is hard to achieve, such as quantum or noisy subproblem solvers. This design balances solution quality, constraint satisfaction, and runtime efficiency within a modular, penalty-driven framework.

\section{Reinforcement Learning-Guided ALM (RL-C-ALM)}
\label{sec:rl-calm}

This section introduces the first major enhancement to the classical framework: the Reinforcement Learning-guided ALM (RL-C-ALM). This method leverages deep reinforcement learning to automate and improve the selection of penalty parameters, which is the most sensitive and manually intensive aspect of the C-ALM. By replacing the reactive, heuristic-based update rules with a proactive, data-driven policy, RL-C-ALM aims to overcome the performance limitations of the classical approach and achieve better solutions more efficiently across a diverse range of problem instances.

\subsection{The Case for Adaptive Penalty Optimization}

The preceding sections described an Augmented Lagrangian Method (ALM) for solving the Capacitated Vehicle Routing Problem (CVRP), in which key constraints, customer visitation and vehicle capacity, are relaxed via dual variables \( \lambda_j, \mu_k \) and associated penalty coefficients \( \rho_j, \sigma_k \). While ALM is theoretically sound, its empirical performance hinges critically on the choice of these penalty parameters.

Traditional heuristics (cf. Equations~\eqref{eq:rho-update}–\eqref{eq:sigma-update}) often require hand-tuning and do not generalize across problem instances of varying size or structure. To overcome these limitations, we propose a reinforcement learning (RL) approach that learns to predict globally optimal penalty parameters, denoted \( \rho_{\text{RL}} \) and \( \sigma_{\text{RL}} \), based on instance-level features.

These RL-predicted values are used to initialize the penalty parameters in Algorithm~\ref{alg:alm-cvrp} (given after the References):
\[
\rho_j^{(0)} \gets \rho_{\text{RL}} \quad \forall j \in C, \qquad
\sigma_k^{(0)} \gets \sigma_{\text{RL}} \quad \forall k \in \{1, \dots, K_{\max}\}.
\]

When penalties are set by the RL agent, the penalty update rules are disabled by fixing the growth rates:
\[
f_\rho = f_\sigma = 1.0.
\]
This ensures penalty coefficients remain constant during the ALM iterations. Meanwhile, the multipliers \( \lambda_j \) and \( \mu_k \) continue to be updated according to Equations~\eqref{eq:lambda-update} and~\eqref{eq:mu-update}.

\subsection{Markov Decision Process (MDP) Formulation}

The learning task is modeled as a Markov Decision Process (MDP), defined by the tuple \( (\mathcal{S}, \mathcal{A}, \mathcal{P}, R, \gamma) \):

\paragraph{State Space \( \mathcal{S} \).}

Each state \( s \in \mathbb{R}^{6} \) encodes a CVRP instance via six informative features:
\begin{enumerate}
    \item Number of customers: \( N_c \)
    \item Vehicle capacity: \( Q  \)
    \item Mean customer demand: \( \bar{d} \)
    \item Standard deviation of demand: \( \text{std}(d_i) \)
    \item Total demand: \( D_{\text{total}} = \sum_i d_i \)
    \item Demand-to-capacity ratio: \( D_{\text{total}} / (Q \cdot N_c) \)
\end{enumerate}

\paragraph{Action Space \( \mathcal{A} \).}

The action \( a = [\rho_{\text{pred}}, \sigma_{\text{pred}}] \in \mathbb{R}^2 \) is a continuous vector where:
\[
\rho_{\text{pred}} \in [\rho_{\min}, \rho_{\max}], \quad
\sigma_{\text{pred}} \in [\sigma_{\min}, \sigma_{\max}].
\]
These correspond to penalty parameters for customer visit and capacity constraints, respectively.

\paragraph{Reward Function \( R \).}

The agent receives a scalar reward \( r_t \) after observing the solution produced by ALM with the chosen penalties. The reward incentivizes:
\begin{itemize}
    \item Proximity to the best known solution (BKS),
    \item Self-improvement over previous agent performance,
    \item Efficiency in convergence (penalizing excessive ALM iterations).
\end{itemize}

Let the following be defined:
\begin{itemize}
    \item \( C_{\text{current}} \): Cost of the current ALM solution
    \item \( C_{\text{BKS}} \): Best known cost for the instance
    \item \( C_{\text{agent-best}} \): Best feasible solution cost found so far by the agent
    \item \( I_{\text{ALM}} \): Number of ALM iterations used
    \item \( \epsilon \): Small constant for numerical stability (e.g., \(10^{-6}\))
\end{itemize}

The reward function is:
\[
r_t =
\begin{cases}
\text{InstanceReward}(C_{\text{current}}, C_{\text{BKS}}) + \text{SelfImprovementBonus}(C_{\text{current}}, C_{\text{agent-best}}) - c_{\text{iter}} \cdot I_{\text{ALM}}, & \text{if } S_{\text{ALM}} \text{ is feasible} \\
R_{\text{infeasible}}, & \text{otherwise}
\end{cases}
\]

Where:
\[
\text{InstanceReward} = \log\left(\frac{C_{\text{BKS}} + \epsilon}{C_{\text{current}} + \epsilon}\right), \quad
\text{SelfImprovementBonus} = \log\left(\frac{C_{\text{agent-best}} + \epsilon}{C_{\text{current}} + \epsilon}\right),
\]
and \( c_{\text{iter}} > 0 \) is a hyperparameter controlling the penalty for long ALM runs.

If ALM fails to return a feasible solution, a fixed negative reward \( R_{\text{infeasible}} \ll 0 \) is issued. This encourages the agent to avoid overly aggressive or ineffective penalty settings.

\paragraph{InstanceReward (Relative to BKS).}

This component evaluates the current solution's quality relative to the instance-specific Best Known Solution (BKS). It provides a baseline reward for matching the BKS and additional incentives for exceeding it, while penalizing deviation otherwise.

\[
\text{InstanceReward} =
\begin{cases}
R_{\text{base}} + R_{\text{bonus\_beatBKS}} \cdot \left( \dfrac{C_{\text{BKS}} - C_{\text{current}}}{\max(C_{\text{BKS}}, \epsilon)} \right), & \text{if } C_{\text{current}} \leq C_{\text{BKS}}, \\
R_{\text{base}} \cdot \max\left(0, 1 - k_{\text{gap}} \cdot \left( \dfrac{C_{\text{current}} - C_{\text{BKS}}}{\max(C_{\text{BKS}}, \epsilon)} \right) \right), & \text{otherwise}.
\end{cases}
\]

\begin{itemize}
    \item \( R_{\text{base}} \): Base reward for reaching the BKS.
    \item \( R_{\text{bonus\_beatBKS}} \): Bonus for surpassing the BKS.
    \item \( k_{\text{gap}} \): Scaling factor for penalizing suboptimal solutions.
    \item \( \epsilon \): Numerical stability constant (e.g., \( 10^{-6} \)).
\end{itemize}

\paragraph{SelfImprovementBonus (Agent-Specific Learning Progress).}

This term encourages the agent to outperform its own historical best solution, even when the BKS cannot be reached. It supports global training progress by reinforcing new personal records.

\[
\text{SelfImprovementBonus} =
\begin{cases}
B_{\text{first\_feasible}}, & \text{if } C_{\text{agent\_best}} = \infty, \\
B_{\text{new\_best}} + k_{\text{improve}} \cdot \left( \dfrac{C_{\text{agent\_best}} - C_{\text{current}}}{\max(C_{\text{agent\_best}}, \epsilon)} \right), & \text{if } C_{\text{current}} < C_{\text{agent\_best}}, \\
0, & \text{otherwise}.
\end{cases}
\]

\begin{itemize}
    \item \( B_{\text{first\_feasible}} \): One-time reward for first feasible solution found.
    \item \( B_{\text{new\_best}} \): Reward for improving the agent’s personal best.
    \item \( k_{\text{improve}} \): Scaling factor for improvement bonus.
\end{itemize}

\paragraph{Iteration Penalty.}

To encourage faster convergence within the ALM loop, we apply a cost for longer iteration counts:

\[
\text{Penalty} = c_{\text{iter}} \cdot I_{\text{ALM}},
\]

where \( c_{\text{iter}} \) is a constant coefficient (e.g., 0.5), and \( I_{\text{ALM}} \) is the number of ALM iterations required.

\paragraph{Penalty for Infeasible Solutions.}

Infeasible solutions receive a large negative reward to discourage penalty parameter settings that lead to constraint violations:

\[
r_t = R_{\text{infeasible}}, \quad \text{where } R_{\text{infeasible}} \ll 0 \text{ (e.g., } -200 \text{)}.
\]

\medskip

\noindent
This composite reward structure balances three critical goals:
\begin{enumerate}
    \item \textbf{Instance-Level Benchmarking} (via BKS comparison),
    \item \textbf{Agent-Specific Learning Progress} (via global best tracking),
    \item \textbf{Optimization Efficiency} (via iteration penalties).
\end{enumerate}

It ensures that the RL agent receives meaningful feedback for both feasible solution quality and learning progress, while strongly enforcing feasibility as a prerequisite for high rewards.

\paragraph{Remaining MDP Elements.}

\begin{itemize}
    \item \textbf{Transitions \( \mathcal{P} \):} Each episode corresponds to a single CVRP instance. The environment resets to a new instance at the end of each episode; there is no temporal transition within episodes.
    
    \item \textbf{Discount Factor \( \gamma \):} Set to \( \gamma = 1.0 \), as each episode consists of a single decision step. No credit propagation across time is required.
\end{itemize}

\subsection{Policy Learning with Soft Actor-Critic (SAC)}

To learn a mapping from instance features to penalty parameters, we adopt the Soft Actor-Critic (SAC) algorithm~\cite{Graf2019}, a model-free, off-policy reinforcement learning method for continuous action spaces. The SAC objective encourages both high rewards and policy entropy, balancing exploration and exploitation.

The objective maximized by the policy \( \pi_\phi(a \mid s) \) is:
\begin{equation}
J(\pi) = \sum_{t=0}^{T} \mathbb{E}_{(s_t, a_t) \sim \mathcal{D}} \left[ r(s_t, a_t) + \alpha \, \mathcal{H}(\pi(\cdot \mid s_t)) \right],
\label{eq:sac-objective-rl-section}
\end{equation}
where \( \mathcal{D} \) is a replay buffer, \( \alpha \) is a learnable temperature coefficient, and \( \mathcal{H}(\pi) \) denotes the policy entropy.

\paragraph{Actor Network \( \pi_\phi(a \mid s) \).}

The actor parameterized by \( \phi \) outputs the mean \( \mu_\phi(s) \) and log-standard deviation \( \log \sigma_\phi(s) \) of a Gaussian policy:
\[
a' \sim \pi_\phi(a \mid s) = \tanh(\mathcal{N}(\mu_\phi(s), \sigma_\phi^2(s))) \in [-1, 1]^2.
\]
This output is scaled to the feasible penalty bounds \( [\rho_{\min}, \rho_{\max}] \) and \( [\sigma_{\min}, \sigma_{\max}] \).

The actor loss is:
\begin{equation}
L_{\text{actor}}(\phi) = \mathbb{E}_{s \sim \mathcal{D},\, a' \sim \pi_\phi} \left[
\alpha \log \pi_\phi(a' \mid s) - \min_{i=1,2} Q_{\theta_i}(s, a')
\right].
\end{equation}

\paragraph{Critic Networks \( Q_{\theta_1}, Q_{\theta_2} \).}

Two Q-functions are trained to reduce overestimation bias. Each is updated via:
\begin{equation}
L_{\text{critic}}(\theta_i) = \mathbb{E}_{(s, a, r, s', d) \sim \mathcal{D}} \left[
\left( Q_{\theta_i}(s, a) - y \right)^2
\right],
\end{equation}
where the target \( y \) is computed using target networks \( \bar{\theta}_j \) and a sampled next action \( a'' \sim \pi_\phi(\cdot \mid s') \):
\[
y = r + \gamma (1 - d) \left[
\min_{j=1,2} Q_{\bar{\theta}_j}(s', a'') - \alpha \log \pi_\phi(a'' \mid s')
\right],
\]
with \( d = 1 \) indicating episode termination.

\paragraph{Entropy Temperature \( \alpha \).}

The temperature parameter \( \alpha \) is updated to maintain a target entropy \( \bar{\mathcal{H}} \), encouraging sufficient exploration:
\begin{equation}
L(\alpha) = \mathbb{E}_{a \sim \pi_\phi(\cdot \mid s)} \left[
- \log \alpha \cdot \left( \log \pi_\phi(a \mid s) + \bar{\mathcal{H}} \right)
\right].
\end{equation}

\subsection{Integration into the ALM Framework}

The learned SAC policy \( \pi_\phi \) is queried once per CVRP instance to generate penalty initialization values. These actions \( a = [\rho_{\text{pred}}, \sigma_{\text{pred}}] \) are integrated into ALM as follows:

\begin{enumerate}
    \item \textbf{Penalty Initialization:} (Lines 2--5 in Algorithm~\ref{alg:alm-cvrp})
    \[
    \rho_j^{(0)} \gets \rho_{\text{pred}}, \quad
    \sigma_k^{(0)} \gets \sigma_{\text{pred}} \quad \forall j \in C,\; \forall k \in \{1, \dots, K_{\max}\}.
    \]

    \item \textbf{Fixed Penalties:} (Overrides Lines 29 and 31)
    \[
    f_\rho = f_\sigma = 1.0 \quad \Rightarrow \quad
    \rho_j^{(t+1)} = \rho_j^{(t)}, \quad
    \sigma_k^{(t+1)} = \sigma_k^{(t)}.
    \]

    \item \textbf{Lagrange Multiplier Updates:} (Preserve Lines 29 and 31) \\
    The multipliers \( \lambda_j^{(t)} \) and \( \mu_k^{(t)} \) continue to be updated at each ALM iteration using Equations~\eqref{eq:lambda-update} and~\eqref{eq:mu-update}.
\end{enumerate}

\noindent
This integration represents a significant architectural shift. In C-ALM, the control paradigm is a dynamic, internal feedback loop where penalties are adjusted reactively. In RL-C-ALM, this is replaced by a static, externally configured system where the intelligence is front-loaded into the initial parameter selection. The ALM solver itself becomes a deterministic procedure once initialized by the RL agent. This design decouples the instance-specific strategic decision (made by the RL policy) from the iterative optimization mechanics (performed by the ALM). The success of this approach, as demonstrated by the empirical results, indicates that the agent learns a genuinely effective and generalizable policy for configuring the solver from the outset.

\section{Quantum-Enhanced Augmented Lagrangian Methods}

This section introduces the integration of quantum computing into the ALM framework. Two variants are presented: a baseline Quantum-Enhanced ALM (Q-ALM) and a fully integrated RL-guided version (RL-Q-ALM).

\subsection{The Quantum-Enhanced ALM (Q-ALM)}

The Q-ALM method serves as the baseline for quantum integration. It combines the classical outer loop of the C-ALM, including its heuristic penalty update mechanism, with a quantum subproblem solver. In this framework, the classical ESP solver used for route generation is replaced by a quantum routine that solves the subproblem via a Quadratic Unconstrained Binary Optimization (QUBO) formulation.

\subsubsection{QUBO Formulation}

 At each ALM iteration, the subproblem of generating a route for a subset of customers is mapped to a QUBO model. The objective function of this QUBO is constructed to include the original travel costs between nodes $(c_{ij})$ as well as the rewards derived from the current Lagrange multipliers $(R_v = -\lambda_v^t)$. This ensures that the quantum solver is guided by the same dual information as its classical counterpart.

\subsubsection{Quantum Solver} 

The formulated QUBO is then mapped to an Ising Hamiltonian, and the Variational Quantum Eigensolver (VQE) is used to find its approximate ground state. The VQE is a hybrid quantum-classical algorithm that uses a classical optimizer to train the parameters of a quantum circuit (an "ansatz") to prepare a quantum state that minimizes the energy of the Hamiltonian. For this work, a shallow, hardware-efficient EfficientSU2 ansatz was used to enhance trainability and mitigate the challenges of deep quantum circuits. The bitstring corresponding to the lowest-energy state found by the VQE is then decoded back into a vehicle route. By pairing this quantum solver with the simple heuristic penalty updates of C-ALM, Q-ALM isolates the effect of the quantum component and provides a clear baseline for evaluating its performance.

\subsubsection{Constraint Handling and Subtour Eliminiation}

A key challenge in routing problems is the elimination of invalid subtours. Within the QUBO model, this is handled by introducing additional penalty terms. A one-hot permutation encoding is used, where binary variables enforce that each customer is assigned to a unique position in the tour. The penalty weights for these constraints are set automatically by the quantum software framework (e.g., Qiskit) to a value larger than any possible objective cost, theoretically ensuring that any feasible solution will have a lower energy than any infeasible one.

\subsection{The RL-Guided Quantum-Enhanced ALM (RL-Q-ALM)}

The RL-Q-ALM represents the full synthesis of all the components developed in this work: the overarching ALM framework, the RL-based parameter tuning, and the quantum subproblem solver. This fully hybrid system leverages reinforcement learning to intelligently configure a quantum-powered optimization loop.

Each reinforcement learning episode corresponds to solving a single CVRP instance using the quantum-enhanced ALM. The cycle is as follows:

\begin{enumerate}
    \item \textbf{Environment Reset \& State Encoding.}
    \begin{itemize}
        \item A small-scale CVRP instance \( \mathcal{I}_t \) is sampled, compatible with the quantum hardware or simulator.
        \item Instance-level features are extracted: \( s_t = f(\mathcal{I}_t) \in \mathbb{R}^6 \), representing the RL state.
    \end{itemize}
    
    \item \textbf{Agent Action Selection.}
    \begin{itemize}
        \item The SAC agent samples an action \( a_t = [\hat{\rho}, \hat{\sigma}] \sim \pi_\phi(a \mid s_t) \).
        \item This action is scaled to produce initial penalty values \( \rho_{\text{init}}, \sigma_{\text{init}} \), bounded as:
        \[
        \rho_{\text{init}} \in [\rho_{\min}, \rho_{\max}], \quad \sigma_{\text{init}} \in [\sigma_{\min}, \sigma_{\max}].
        \]
    \end{itemize}

    \item \textbf{ALM Initialization.}
    \begin{itemize}
        \item A new instance of the quantum-enhanced solver is initialized with:
        \[
        \rho_j^{(0)} \gets \rho_{\text{init}}, \quad \sigma_k^{(0)} \gets \sigma_{\text{init}}, \quad \forall j \in C,\, \forall k \in \{1, \dots, K_{\max}\}.
        \]
        \item Penalty growth rates are disabled: \( f_\rho = f_\sigma = 1.0 \).
    \end{itemize}

    \item \textbf{Quantum-Enhanced ALM Execution.}
    \begin{itemize}
        \item \textbf{Outer Loop.} The ALM iterates up to a maximum of \( T \) iterations or until convergence.
        
        \item \textbf{Subproblem Generation.} In each iteration, a subset of customers \( S_{\text{sub}} \subseteq C \) (typically \( |S_{\text{sub}}| = 2 \) or 3) is selected based on current multipliers \( \lambda^t \).
        
        \item \textbf{QUBO Formulation.} Using rewards \( R_v = -\lambda_v^t \), the subproblem is encoded as:
        \[
        \min_{x} \sum_{(u,v)} c_{uv} x_{uv} + \sum_{v \in S_{\text{sub}}} R_v x_v + \text{penalty terms}.
        \]
        This forms a TSP-QUBO over the subgraph \( S_{\text{sub}} \cup \{0\} \).

        \item \textbf{QUBO Solving via VQE.} The QUBO is mapped to an Ising Hamiltonian and solved with VQE. The returned bitstring is decoded into a route.

        \item \textbf{Route Construction.} This process is repeated until a complete set of routes \( X^{(t+1)} \) is built.

        \item \textbf{Dual Updates.} Same as \emph{C-ALM} for multiplier/penalty updates (Section~\ref{sec:calm}).
    \end{itemize}
    
    \item \textbf{Reward computation} Use the same definition and weights as Section~\ref{sec:rl-calm}; no RL-specific changes are introduced by the quantum backend.
    
    \item \textbf{Policy update} Same SAC updates as Section~\ref{sec:rl-calm}; we only change the downstream solver (quantum vs.\ classical), not the learning algorithm.
\end{enumerate}

\subsubsection{Learning Objective and Environment Design}

The agent’s objective is to learn a policy \( \pi_\phi(a_t \mid s_t) \) that maps features of a given CVRP instance \( s_t \in \mathbb{R}^6 \) to penalty parameters \( a_t = (\rho_{\text{init}}, \sigma_{\text{init}}) \in \mathbb{R}^2 \), which initialize the Augmented Lagrangian Method (ALM). These penalty values critically influence the convergence speed, feasibility, and solution quality of the resulting optimization process.

While prior constrained optimization environments often involve deterministic solvers, our setup integrates a quantum-inspired backend: the ALM subproblems are solved via a QUBO-based formulation using the Variational Quantum Eigensolver (VQE). Although simulations are noiseless, this quantum variational process is inherently non-convex, approximate, and sensitive to initial conditions and ansatz depth, making convergence behavior difficult to predict a priori.

This gives rise to several reinforcement learning challenges:
\begin{itemize}
    \item \textbf{Sample Efficiency.} Since evaluating a full ALM run per action is costly, the agent must learn efficiently from sparse feedback.
    \item \textbf{Reward Smoothness.} Even without quantum noise, the VQE solver can produce non-monotonic improvements, so the reward signal must be shaped to reflect feasibility, optimality, and convergence cost.
    \item \textbf{Cross-Instance Generalization.} The policy must generalize across CVRP instances of varying scale, demand distributions, and depot-customer geometries.
\end{itemize}

To address these challenges, we implement a custom Gym-style environment, which:
\begin{itemize}
    \item Encodes CVRP instance statistics into a fixed-length feature vector;
    \item Interfaces seamlessly with quantum-enhanced ALM solvers for route generation;
    \item Returns scalar rewards that combine solution feasibility, optimality gap (vs. BKS), and convergence cost (see Section~4).
\end{itemize}

This design allows the agent to be trained in a structured, reproducible, and computationally efficient environment, while still capturing the critical dynamics of hybrid quantum-classical optimization workflows.

\vspace{0.5em}
\noindent
\textbf{Comparison to Prior Work.} 
Ayanzadeh et al.~\cite{Ayanzadeh2020} proposed reinforcement quantum annealing, where penalty weights in QUBO formulations are iteratively adjusted based on feedback from a D-Wave quantum annealer. However, their approach lacks a structured decomposition or classical optimization framework. In contrast, RL-Q-ALM uses RL to predict global penalty parameters which are fed into a classical augmented Lagrangian loop, with route subproblems solved using a gate-based variational quantum solver (VQE). This structured hybrid enables better interpretability and scalability, particularly for solving CVRP-like problems with combinatorial constraints.

\section{Empirical Evaluation}
\label{sec:empirical}

This section presents a comparative empirical analysis of four solver paradigms for the CVRP.
The evaluation spans three classes of CVRP instances to capture a range of practical and quantum-relevant complexities: (1) small custom-generated instances suitable for QUBO encoding and quantum simulation, (2) medium-sized synthetic instances for evaluating scalability and classical solver behavior, and (3) real-world benchmark instances from CVRPLIB for compatibility and reproducibility.

Key metrics include:
\begin{itemize}
    \item \textbf{Total Route Cost:} Sum of edge weights across all routes.
    \item \textbf{Optimality Gap:} Deviation from the Best Known Solution (BKS), when available.
    \item \textbf{Feasibility:} Percentage of instances where all ALM constraints are satisfied.
    \item \textbf{Convergence Behavior:} Runtime.
\end{itemize}

Our findings demonstrate that RL-based tuning improves ALM convergence and feasibility in both classical and quantum settings, and that QUBO-based subproblem solving via VQE is feasible for small-scale CVRP instances under simulation.

\subsection{Benchmark Instance Classes}
\label{subsec:benchmark_instances}

To evaluate cross-method performance and generalization, we curate three distinct datasets:

\subsubsection{Small Custom-Generated Instances}

We generate approximately 50 synthetic instances tailored for quantum feasibility. These instances support direct comparison across all four solver variants.

\begin{itemize}
    \item \textbf{Customers:} \( N_c \in [5, 8] \)
    \item \textbf{Coordinates:} Sampled from a Gaussian distribution centered at the depot (fixed at grid center), with support in \( [0, 25]^2 \); spread controlled by a standard deviation scaling factor \( = 0.1 \)
    \item \textbf{Capacity:} 200 units (fixed)
    \item \textbf{Demands:} Uniformly sampled integers in \( [10, 40] \)
\end{itemize}

These instances are small enough to allow QUBO encoding of subproblems and quantum simulation via VQE. A subset is reserved for testing generalization of the learned RL policy.

\subsubsection{Medium-Scale Synthetic Instances}

A corpus of 35 larger instances is used to evaluate classical scalability:

\textbf{Instance Profiles:} We define two representative instance profiles to capture varying problem scales and constraint structures:
    \begin{itemize}
        \item \textsc{AB-Like}: Number of customers $n \in [25, 40]$, vehicle capacity $Q = 100$, customer demands $d_i \in [5, 25]$, and coordinate range bounded by $[0, 100]^2$.
        \item \textsc{E-Small-Like}: Number of customers $n \in [10, 25]$, vehicle capacity $Q \in [4000, 7000]$, customer demands $d_i \in [100, 1500]$.

    \item \textbf{Spatial Distributions:} Customer locations are generated under three geometric configurations: \emph{uniformly random}, \emph{clustered}, and \emph{skewed}. The depot is placed either at the geometric center of the region or with a deliberate spatial offset to evaluate algorithm robustness under varying service geometries.

    \item \textbf{Distance Metric:} All distances are computed using the Euclidean metric ($\ell_2$ norm), consistent with the \textsc{EUC\_2D} standard in vehicle routing benchmarks.
\end{itemize}

Due to circuit width constraints, quantum approaches are not applied here. This dataset is used solely to assess the scalability of C-ALM and RL-C-ALM.

\subsubsection{Standard CVRPLIB Benchmarks}

To ensure compatibility with existing literature, we evaluate on 14 benchmark instances from the CVRPLIB suite \cite{Uchoa2017} , including:
\begin{itemize}
    \item \texttt{E-n22-k4}, \texttt{A-n32-k5}, \texttt{B-n34-k5}, etc.
\end{itemize}

These benchmarks are used to:
\begin{itemize}
    \item Compare classical solvers against best known solutions (BKS)
    \item Quantify optimality gap and feasibility rates
    \item Establish practical relevance and generalization across diverse instance families
\end{itemize}

This multi-tiered evaluation provides insight into both theoretical feasibility (small instances with quantum support) and real-world performance (larger synthetic and benchmark instances).

\section{Experimental Results}
\label{sec:experiments}

We conduct a comparative empirical evaluation of the four solver variants introduced in this study:

\begin{itemize}
    \item \textbf{C-ALM:} Classical Augmented Lagrangian Method with heuristic penalty updates.
    \item \textbf{RL-C-ALM:} Classical ALM with instance-adaptive penalty initialization learned via Soft Actor-Critic (SAC).
    \item \textbf{Q-ALM:} Quantum-augmented ALM where routing subproblems are solved via QUBO-based VQE.
    \item \textbf{RL-Q-ALM:} Fully hybrid framework combining quantum subproblem solvers with RL-driven penalty tuning.
\end{itemize}

Evaluations were performed on (i) small quantum-tractable CVRP instances, and (ii) medium-scale synthetic instances to assess classical scalability. Performance is measured by feasibility rate, optimality rate, and average runtime. Results are summarized in Table~\ref{tab:aggregate}.

For the variational form, we employed a hardware-efficient ansatz, specifically, the \texttt{EfficientSU2} circuit from the Qiskit circuit library~\cite{Qiskit}. For quantum simulations, VQE was selected over QAOA primarily due to its greater flexibility in ansatz design, with VQE, one can carefully control the degree of entanglement and circuit depth by tailoring the parameterized quantum circuit to the problem and hardware constraints. In contrast, QAOA's ansatz is inherently problem-dependent, derived directly from the cost Hamiltonian, which can lead to highly complex circuits even at depth one for non-trivial Ising formulations. For certain problems, simulating even a single-layer QAOA circuit becomes computationally infeasible due to the exponential growth in gate complexity.

Acknowledging the challenges associated with deep quantum circuits, including noise accumulation and the risk of barren plateaus, we deliberately constrained the ansatz to a shallow depth by setting the repetition parameter to one. This conservative design choice was motivated by the need to enhance the trainability of the variational parameters, a known sensitivity in variational quantum eigensolver (VQE) algorithms.

\subsection{Aggregate Results on Small and Synthetic Instances}

\begin{table}[ht]
\centering
\caption{Aggregate performance on \textit{light} (50 instances, $N_c \in [5,8]$) and \textit{synthetic} (35 instances, $N_c \in [10,40]$) CVRP datasets. 
Feasibility is the percentage of instances where a method found a constraint-satisfying solution. 
Optimality indicates the percentage of cases where a method matched or outperformed all others (based on feasible cost). 
Quantum methods were only evaluated on small instances due to qubit and simulation limitations.}
\label{tab:aggregate}
\renewcommand{\arraystretch}{1.2}
\begin{tabular}{l|ccc|ccc|ccc|ccc}
\toprule
\textbf{Instance Set} & \multicolumn{3}{c|}{\textbf{C-ALM}} & \multicolumn{3}{c|}{\textbf{RL-C-ALM}} & \multicolumn{3}{c|}{\textbf{Q-ALM}} & \multicolumn{3}{c}{\textbf{RL-Q-ALM}} \\
                      & Feas. & Opt. & Time & Feas. & Opt. & Time & Feas. & Opt. & Time & Feas. & Opt. & Time \\
                      & (\%)  & (\%) & (s)  & (\%)  & (\%) & (s)  & (\%)  & (\%) & (s)  & (\%)  & (\%) & (s) \\
\midrule
\textit{light}        & 100.0 & 98.0 & 0.84 & 100.0 & 98.0 & 0.79 & 100.0 & 92.0 & 211.20 & 100.0 & 98.0 & 166.80 \\
\textit{synthetic}    & 98.0  & 70.0 & 868.5 & 98.0 & 78.0 & 775.68 & -- & -- & -- & -- & -- & -- \\
\bottomrule
\end{tabular}
\end{table}

\subsubsection*{Small Custom Instances (\textit{light})}

These are deliberately small CVRP problems ($N_c \in [5, 8]$) constructed to permit full QUBO encoding and tractable simulation with VQE. Results highlight the relative strengths of classical and quantum solvers under fixed-capacity, short-horizon settings.

\begin{itemize}
    \item \textbf{Feasibility:} All methods achieved 100\% feasibility on these small instances, confirming the effectiveness of ALM-style constraint handling even with approximate subproblem solvers.
    \item \textbf{Runtime:} Classical methods (C-ALM and RL-C-ALM) were significantly faster, with average runtimes under 1 second. In contrast, quantum methods required several minutes per instance due to VQE overhead: Q-ALM averaged 211.2s, while RL-Q-ALM improved upon this with 166.8s.
    \item \textbf{Solution Quality:} RL-Q-ALM achieved 98.0\% optimality—matching both classical variants—while Q-ALM lagged slightly at 92.0\%. These results suggest that RL tuning improves both convergence and robustness in the quantum setting.
\end{itemize}

\subsubsection*{Synthetic Medium-Sized Instances (\textit{synthetic})}

These 35 instances ($N_c \in [10, 40]$) are representative of realistic logistics settings. Quantum solvers were excluded from this benchmark due to the prohibitive qubit count required for VQE-based QUBO simulation.

\begin{itemize}
    \item \textbf{Feasibility:} Both C-ALM and RL-C-ALM achieved 98.0\% feasibility, indicating the continued viability of the ALM framework at moderate scales.
    \item \textbf{Optimality:} RL-C-ALM outperformed C-ALM in 78.0\% of instances versus 70.0\%, showing that learned penalty initialization can enhance solution quality on harder problems.
    \item \textbf{Runtime:} RL-C-ALM also ran faster on average (775.68s vs 868.5s), likely due to faster convergence and more stable ALM trajectories.
\end{itemize}

\subsubsection*{Interpretation}

From these aggregate results, we draw three key conclusions:
\begin{enumerate}
    \item RL tuning improves classical ALM both in solution quality and convergence time, especially as instance complexity increases.
    \item Quantum-enhanced ALM is viable on small instances and, with RL, can match classical performance in terms of optimality.
    \item The main limitation of quantum approaches is computational cost—even under ideal (noiseless) simulation—necessitating hybrid strategies and scalable encoding improvements for future deployment.
\end{enumerate}

\subsection{Detailed Benchmark Performance}
\label{subsec:benchmark_performance}

To further assess solver robustness and real-world applicability, we evaluated all methods on a curated subset of standard CVRP benchmark instances from CVRPLIB. Table~\ref{tab:flat_gap_runtime} reports the optimality gap (relative to Best Known Solutions, or BKS) and average runtime per instance for each solver.

\begin{table*}[t]
\caption{Optimality gap (\%) and runtime (s) for each method across instances.}
\label{tab:flat_gap_runtime}
\centering

\newcolumntype{Y}{>{\centering\arraybackslash}X}

\begin{tabularx}{\textwidth}{l *{4}{YY}}
\toprule
\textbf{Instance} 
& \multicolumn{2}{c}{\textbf{C-ALM}} 
& \multicolumn{2}{c}{\textbf{RL-C-ALM}} 
& \multicolumn{2}{c}{\textbf{Q-ALM}} 
& \multicolumn{2}{c}{\textbf{RL-Q-ALM}} \\
\cmidrule(lr){2-3} \cmidrule(lr){4-5} \cmidrule(lr){6-7} \cmidrule(lr){8-9}
& Gap (\%) & Time (s) 
& Gap (\%) & Time (s) 
& Gap (\%) & Time (s) 
& Gap (\%) & Time (s) \\
\midrule
\textbf{E-n13-k4}  & 0.00 & 7.48     & 0.00 & 6.68 & 14.17 & 3354.73 & 4.04 & 2875.63 \\
\textbf{E-n22-k4}  & 4.53 & 435.96   & 3.20 & 355.64 & 22.93 & 9216.58 & 20.53 & 8152.59 \\
\textbf{E-n23-k3}  & 0.00 & 1041.59  & 1.40 & 983.59 & 9.66 & 9955.44 & 10.54 & 8710.78 \\
\textbf{E-n30-k3}  & 7.86 & 2136.31  & 4.11 & 1816.95 & 17.60 & 11835.36 & 14.23 & 10460.11 \\
\textbf{E-n31-k7}  & 8.70 & 6980.96  & 5.01 & 6179.24 & 20.31 & 13914.72 & 19.26 & 12215.95 \\
\textbf{B-n31-k5}  & 4.01 & 2582.80  & 3.12 & 2243.86 & 17.11 & 12830.76 & 19.64 & 11244.06 \\
\textbf{B-n34-k5}  & 8.62 & 8347.20  & 4.82 & 7372.53 & -- & -- & -- & -- \\
\textbf{B-n35-k5}  & 9.52 & 15711.69 & 6.49 & 13845.28 & -- & -- & -- & -- \\
\textbf{A-n32-k5}  & 17.09 & 10913.16 & 16.58 & 9626.58 & 19.64 & 16873.20 & 18.36 & 14891.41 \\
\textbf{A-n33-k5}  & 16.64 & 8525.33  & 11.95 & 7503.29 & -- & -- & -- & -- \\
\textbf{A-n33-k6}  & 14.42 & 15048.15 & 4.98 & 13202.37 & -- & -- & -- & -- \\
\textbf{A-n34-k5}  & 13.49 & 10424.14 & 10.79 & 9142.24 & -- & -- & -- & -- \\
\textbf{A-n36-k5}  & 25.03 & 16108.41 & 23.77 & 14173.40 & -- & -- & -- & -- \\
\textbf{A-n37-k5}  & 14.20 & 16654.23 & 10.76 & 14675.72 & -- & -- & -- & -- \\
\bottomrule
\end{tabularx}
\end{table*}

\begin{figure}[ht]
  \centering
  \begin{subfigure}[b]{0.48\textwidth}
    \includegraphics[width=\textwidth]{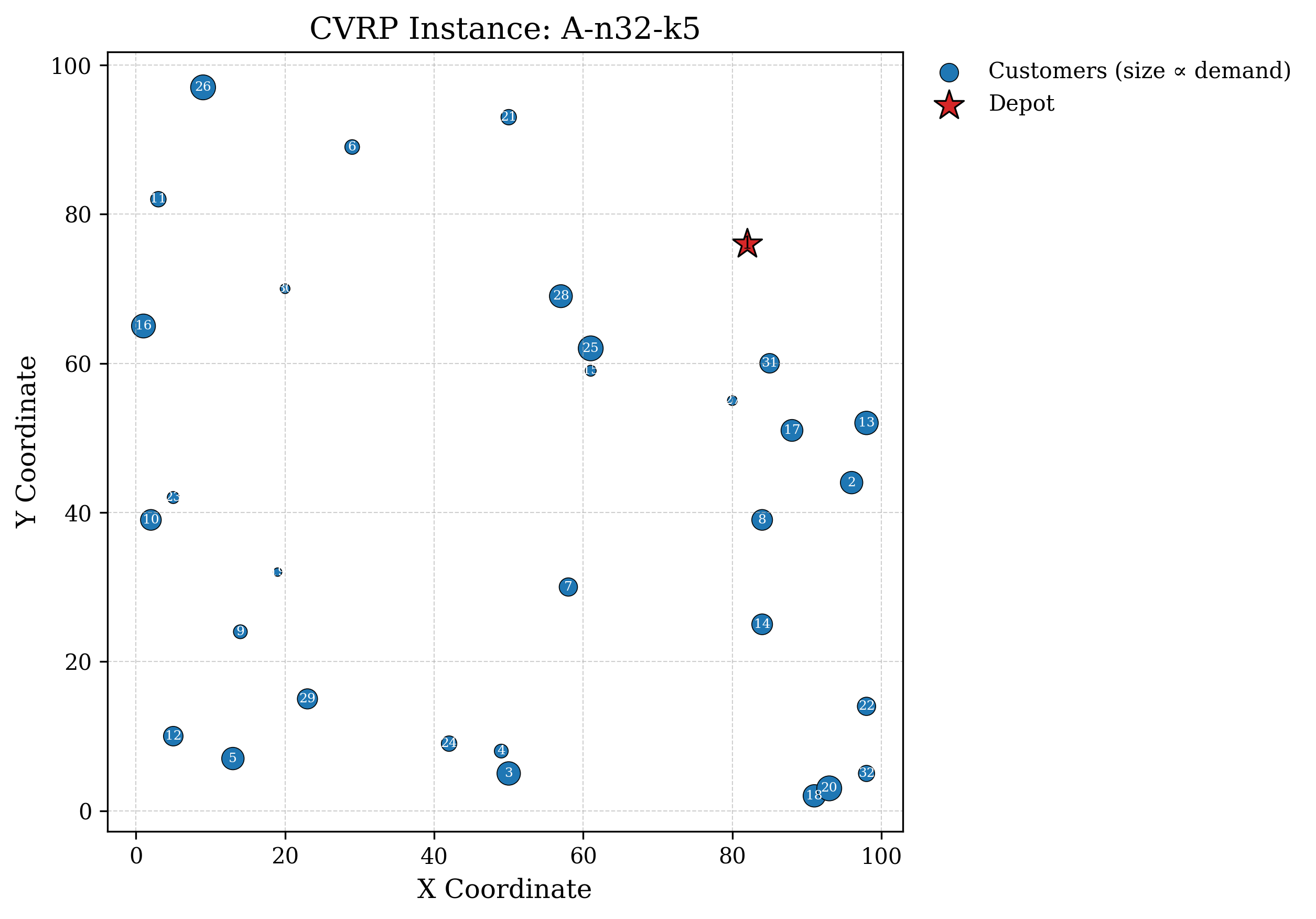}
    \caption{Problem Instance}
    \label{fig:instance}
  \end{subfigure}
  \hfill
  \begin{subfigure}[b]{0.48\textwidth}
    \includegraphics[width=\textwidth]{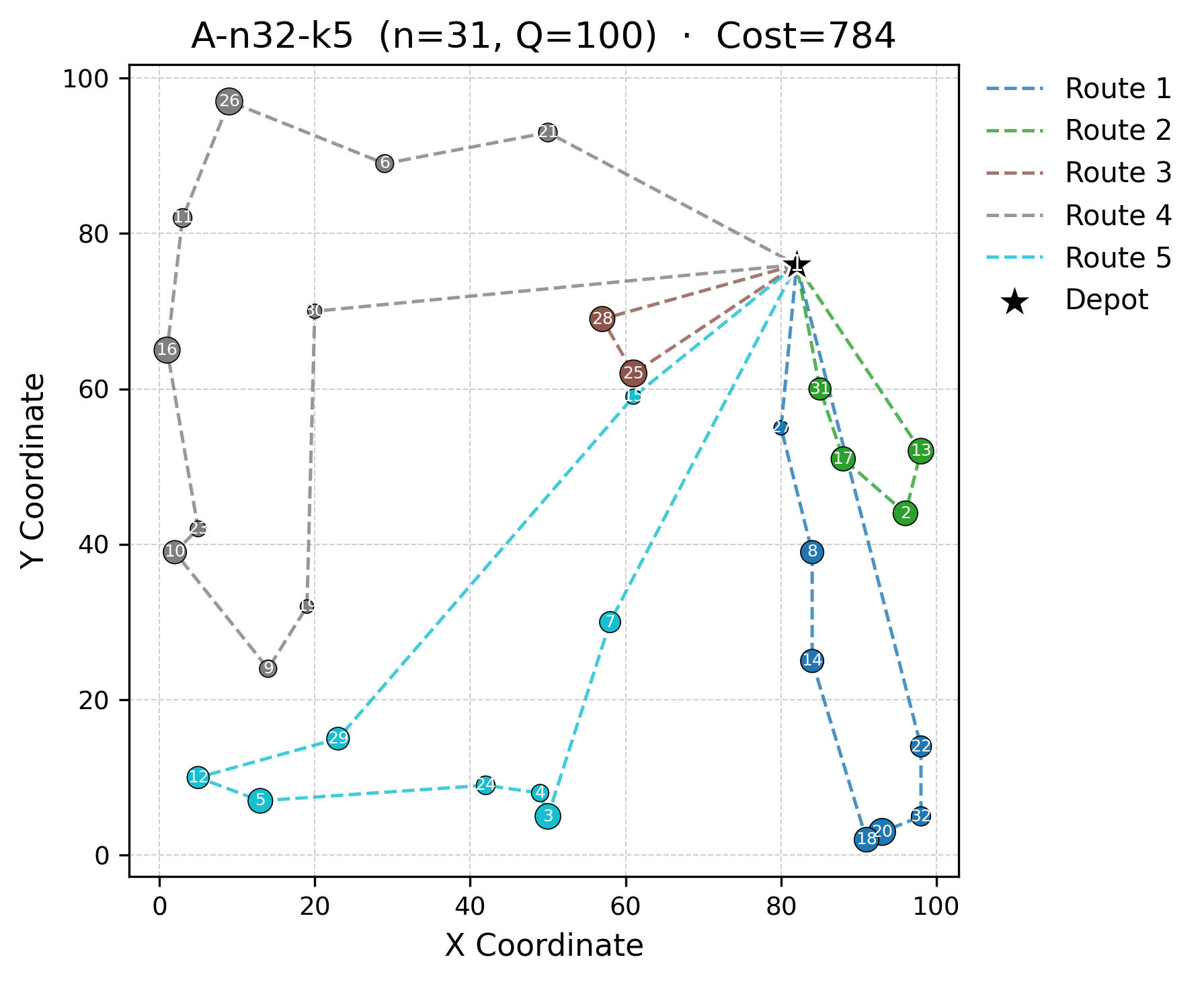}
    \caption{Known Optimal Solution}
    \label{fig:optimal-solution}
  \end{subfigure}
  \caption{Visualization of the CVRP instance A-n32-k5. 
(a) Problem layout showing depot ($\star$) and customer locations (size $\propto$ demand). 
(b) Optimal solution with 5 routes and total cost of 784.}

  \label{fig:rl-alm-comparison1}
\end{figure}

\begin{figure}[ht]
  \centering
  \begin{subfigure}[b]{0.48\textwidth}
    \includegraphics[width=\textwidth]{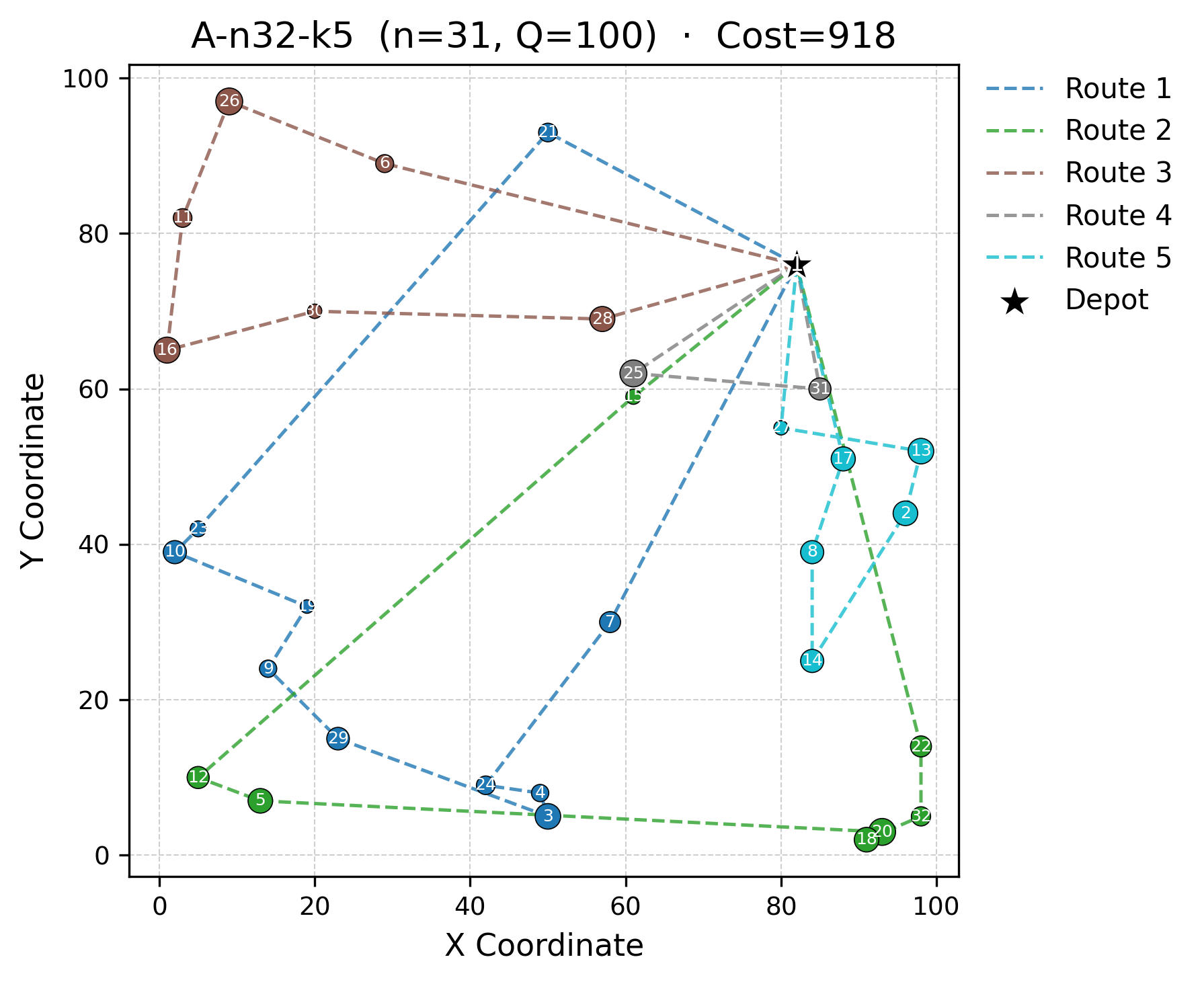}
    \caption{C-ALM Solution}
    \label{fig:c-alm-solution}
  \end{subfigure}
  \hfill
  \begin{subfigure}[b]{0.48\textwidth}
    \includegraphics[width=\textwidth]{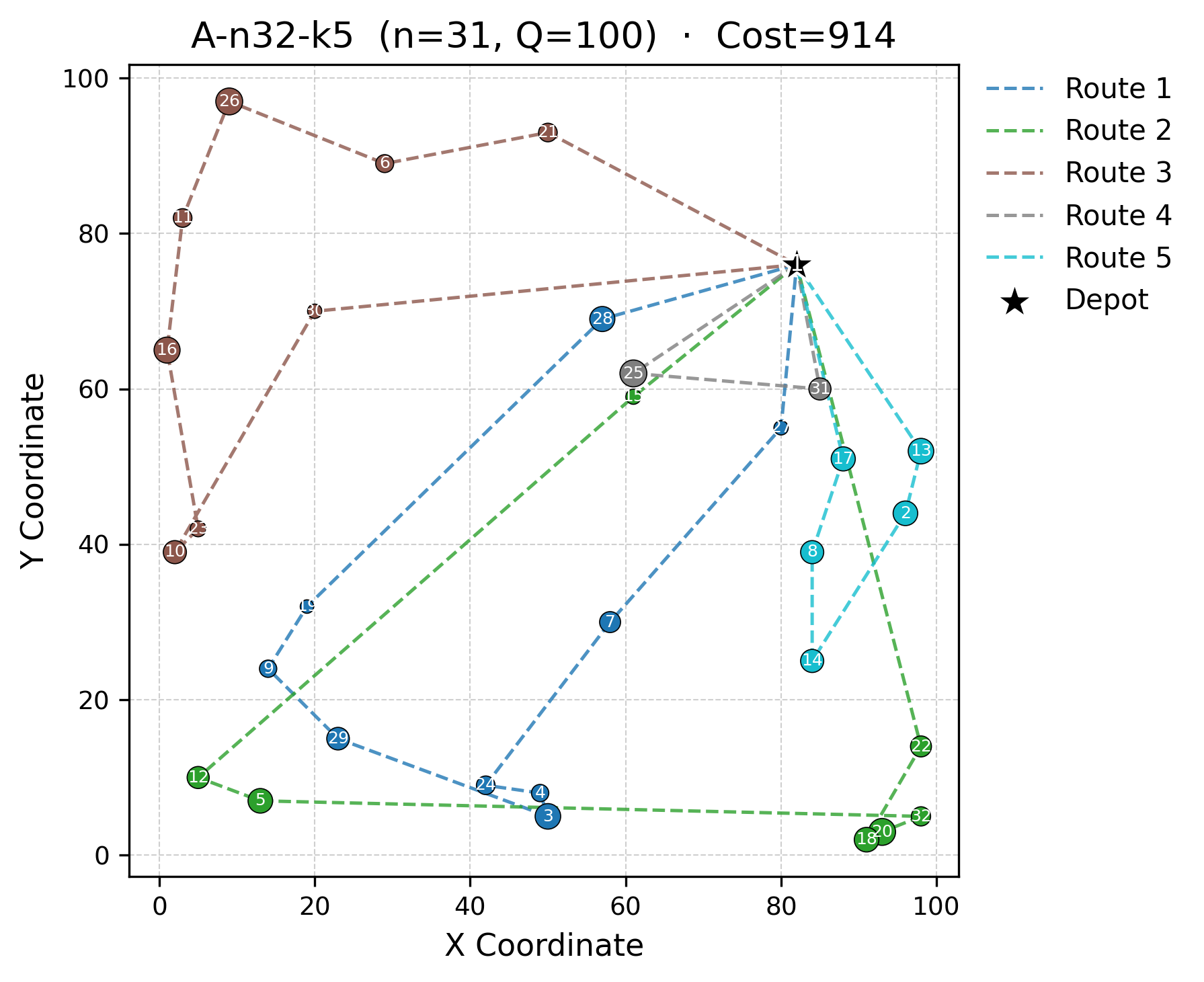}
    \caption{RL-C-ALM Solution}
    \label{fig:rl-c-alm-solution}
  \end{subfigure}
  \caption{Classical ALM Solvers on A-n32-k5. 
(a) C-ALM yields a cost of 918. 
(b) RL-C-ALM improves the solution to 914 through RL-guided penalty tuning.}

  \label{fig:rl-alm-comparison2}
\end{figure}

\begin{figure}[ht]
  \centering
  \begin{subfigure}[b]{0.48\textwidth}
    \includegraphics[width=\textwidth]{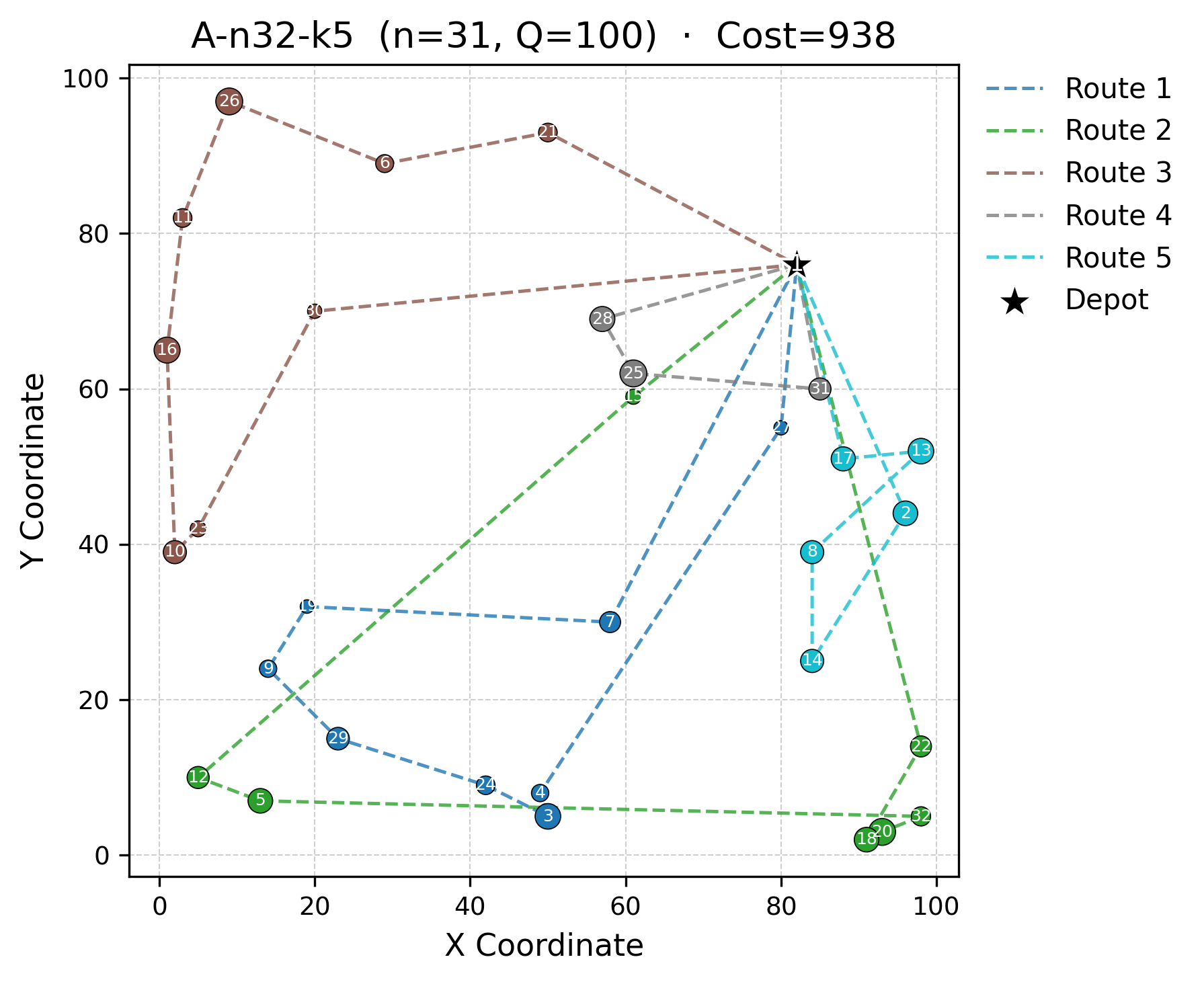}
    \caption{Q-ALM Solution}
    \label{fig:q-alm-solution}
  \end{subfigure}
  \hfill
  \begin{subfigure}[b]{0.48\textwidth}
    \includegraphics[width=\textwidth]{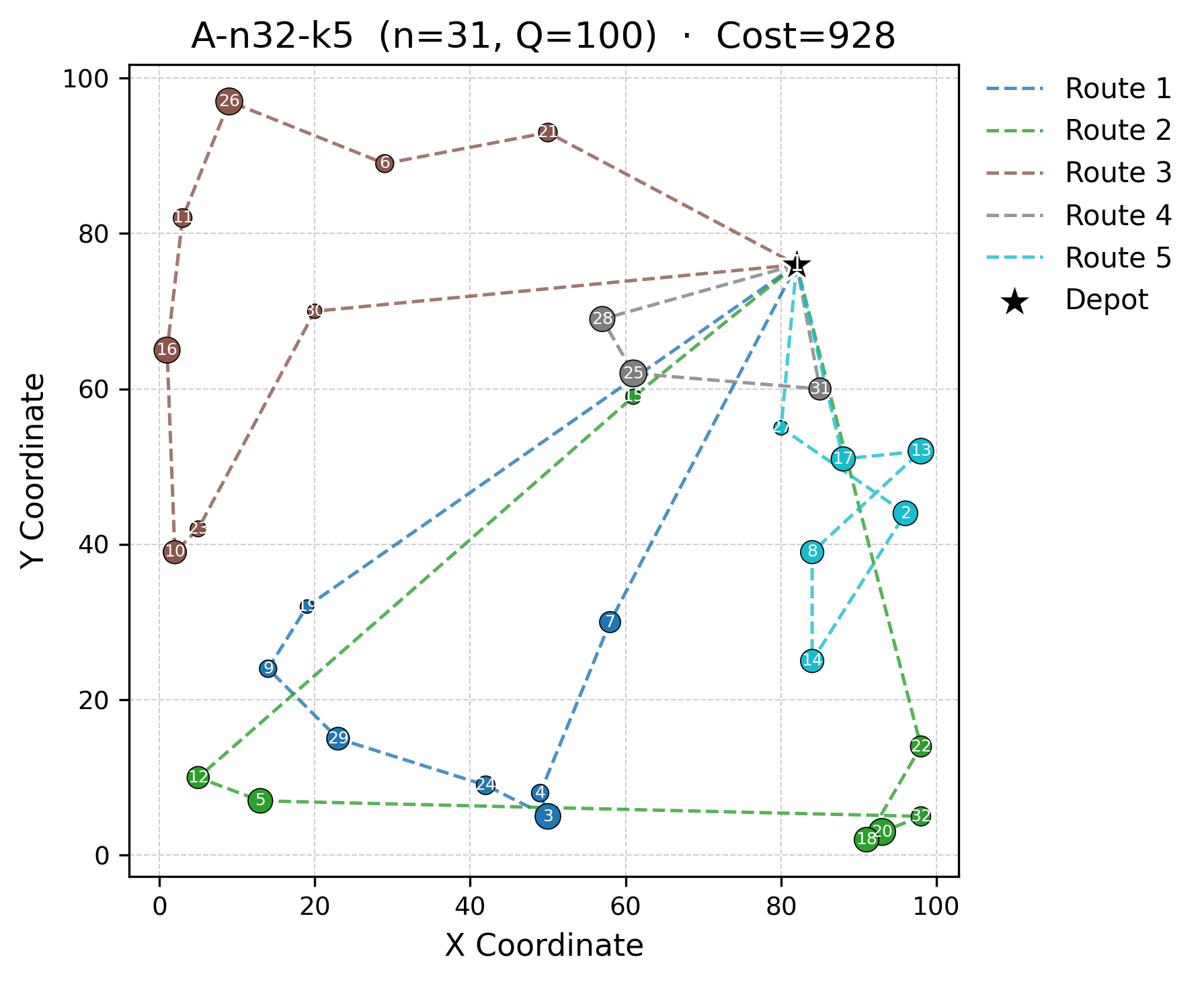}
    \caption{RL-Q-ALM Solution}
    \label{fig:rl-q-alm-solution}
  \end{subfigure}
  \caption{Quantum-Augmented ALM Solvers on A-n32-k5. 
(a) Q-ALM produces a solution with cost 938. 
(b) RL-Q-ALM improves this to 928 by learning better global penalty settings.}

  \label{fig:rl-alm-comparison3}
\end{figure}

To complement the tabulated results, Figure~\ref{fig:instance} shows the visual layout of the CVRP instance A-n32-k5, while Figure~\ref{fig:optimal-solution} displays the known optimal solution. Figures~\ref{fig:c-alm-solution}–\ref{fig:rl-q-alm-solution} compare the solutions generated by all four methods. RL-C-ALM achieves a lower cost than C-ALM by optimizing route balance, while RL-Q-ALM slightly improves the compactness and cost over Q-ALM despite the quantum circuit constraints. These visualizations reinforce the quantitative findings of Table~\ref{tab:flat_gap_runtime}, particularly in showing the spatial efficiency of RL-guided solvers.

\paragraph{Feasibility and Coverage.}
C-ALM and RL-C-ALM successfully produced feasible solutions for all 14 benchmark problems. In contrast, quantum solvers (Q-ALM and RL-Q-ALM) failed to solve 7 of 14 instances due to excessive size, infeasibility under limited qubit budgets, or failure to converge during QUBO decoding.

\paragraph{Classical Solver Comparison.}
RL-C-ALM consistently outperformed C-ALM in solution quality:
\begin{itemize}
    \item \textbf{E-n22-k4:} RL-C-ALM reduced the optimality gap from 4.53\% to 3.20\%.
    \item \textbf{A-n33-k6:} Gap improved from 14.42\% (C-ALM) to 4.98\% (RL-C-ALM).
    \item \textbf{A-n33-k5:} Gap improved from 16.64\% to 11.95\%, while runtime dropped by 1000 seconds.
\end{itemize}

These results highlight the benefits of RL-based penalty initialization in improving both convergence speed and final solution quality across a wide range of instances.

\paragraph{Quantum-Augmented Solvers.}
On small-to-medium instances where QUBO encoding was possible, Q-ALM and RL-Q-ALM yielded weaker solutions than their classical counterparts:
\begin{itemize}
    \item \textbf{E-n13-k4:} Q-ALM had a 14.17\% gap, reduced to 4.04\% with RL-Q-ALM, but still worse than classical methods (0.00\%).
    \item \textbf{E-n23-k3:} Q-ALM and RL-Q-ALM had 9–10\% gaps, compared to 0–1.4\% classically.
\end{itemize}
Despite these drawbacks, RL tuning consistently improved quantum performance (e.g., E-n30-k3: 17.60\% $\rightarrow$ 14.23\%).

\paragraph{Runtime Overhead of Quantum Methods.}
Quantum solvers incurred significantly longer runtimes, with Q-ALM and RL-Q-ALM often requiring $10\times$ to $100\times$ the time of classical counterparts:
\begin{itemize}
    \item \textbf{E-n13-k4:} C-ALM required 7.48s vs. Q-ALM’s 3354.73s.
    \item \textbf{E-n22-k4:} RL-Q-ALM took 8152.59s vs. RL-C-ALM’s 355.64s.
\end{itemize}
These costs stem from iterative QUBO formulation, VQE circuit execution (even in simulation), and bitstring decoding. This makes quantum methods currently impractical for large CVRP instances, despite their conceptual feasibility.

\paragraph{Summary.}
\begin{itemize}
    \item RL-C-ALM outperformed C-ALM across most benchmark instances in both gap and time.
    \item RL-Q-ALM modestly improved Q-ALM’s feasibility and gap when successful.
    \item Classical solvers remain dominant on standard benchmarks due to current quantum limitations, though RL-Q-ALM offers a promising hybrid direction for small, tractable instances.
\end{itemize}

\subsection{Overall Summary of Findings}

Synthesizing the results from Tables~\ref{tab:aggregate} and~\ref{tab:flat_gap_runtime}, we draw the following conclusions:

\begin{itemize}
    \item \textbf{Classical Methods (C-ALM and RL-C-ALM)} are robust across all tested CVRP instances. 
    \begin{itemize}
        \item RL-C-ALM consistently improves upon C-ALM in terms of optimality gap, especially on medium and hard instances, without incurring additional runtime overhead.
        \item Both methods maintain high feasibility (98–100\%) across synthetic and benchmark datasets.
    \end{itemize}
    
    \item \textbf{Quantum-Augmented Methods (Q-ALM and RL-Q-ALM)} show limited applicability under current hardware constraints.
    \begin{itemize}
        \item On small CVRP instances, RL-Q-ALM achieves solution quality comparable to classical methods, and improves over Q-ALM in both optimality gap and runtime.
        \item However, on larger or more complex instances, quantum solvers fail to return feasible solutions due to limitations in QUBO size, circuit depth, or decoding reliability.
        \item Quantum runtimes are significantly higher—often by two orders of magnitude—even on small instances simulated without noise.
    \end{itemize}
    
    \item \textbf{Scalability Limitations:} Current quantum techniques do not scale to CVRP instances with more than $\sim$8 customers, due to qubit count, limited expressibility of the QUBO mapping, and exponential simulation overheads.

    \item \textbf{Impact of RL:} Reinforcement learning consistently enhances ALM performance, both classically and quantumly, by improving convergence and optimality. However, on especially hard instances, RL-guided penalties can sometimes reduce feasibility by encouraging overly aggressive optimization.

\end{itemize}

\noindent These findings reinforce the practical superiority of classical and RL-enhanced solvers for CVRP in the near term. Quantum optimization remains an important long-term direction, but realizing competitive performance on real-world combinatorial problems will require substantial advancements in quantum hardware (e.g., more qubits, better noise handling) and improved QUBO encodings with scalable constraint enforcement mechanisms.

\section{Future Research Directions}

While this work demonstrates the feasibility and potential of integrating reinforcement learning with quantum-augmented optimization for the Capacitated Vehicle Routing Problem (CVRP), several important avenues remain open for future exploration:

\subsection{Quantum Scalability and Hardware Deployment}
Currently, all quantum subproblems were solved on simulators due to the limitations of near-term quantum hardware. As quantum processors advance—with improved coherence times, lower noise rates, and higher qubit counts—it will become possible to test this framework on real quantum devices. Future research should explore:
\begin{itemize}
    \item Empirical performance on noisy intermediate-scale quantum (NISQ) hardware.
    \item Error mitigation techniques tailored to QUBO-based routing problems.
    \item Scalability under hardware-efficient ansätze and qubit reuse strategies.
\end{itemize}

\subsection{Constraint Generalization and Encoding Efficiency}
Our current framework relaxes visit and capacity constraints using an augmented Lagrangian method, with feasibility guided by multiplier updates and penalties. However, subtour elimination is handled differently in the classical and quantum subproblem solvers, reflecting a trade-off between structural guarantees and encoding flexibility. Future directions for improving constraint expressiveness and quantum encoding efficiency include:
\begin{itemize}
    \item Designing QUBO encodings that natively incorporate routing constraints such as subtour elimination, precedence, and time windows.
    \item Applying QUBO compression techniques or diagonal encodings to reduce qubit requirements while preserving constraint fidelity.
    \item Leveraging symmetry-aware or permutation-equivariant architectures to improve generalization and reduce redundant solution space exploration.
\end{itemize}

\subsection{Reinforcement Learning Generalization and Robustness}
Our reinforcement learning agent currently learns global penalty parameters for individual CVRP instances. To further improve generalization:
\begin{itemize}
    \item Explore curriculum learning or meta-RL to enable scalable transfer to larger or structurally different CVRP instances.
    \item Incorporate graph neural networks (GNNs) to encode richer structural features of the CVRP instance graph.
    \item Investigate multi-objective reward functions that balance cost, feasibility, and runtime more effectively.
\end{itemize}

\subsection{Hybrid Optimization Workflow Integration}
A long-term goal is to create a hybrid optimization pipeline that dynamically integrates quantum solvers into classical frameworks. Future research could:
\begin{itemize}
    \item Use RL to not only tune penalties but also to decide \emph{when} and \emph{where} to invoke quantum solvers during optimization.
    \item Combine classical heuristics (e.g., Tabu Search, Local Branching) with quantum subroutines as modular components.
    \item Benchmark across broader problem domains (e.g., VRPTW, multi-depot VRP) to evaluate framework generality.
\end{itemize}

\subsection{Theoretical Analysis of RL-Quantum Convergence}
Finally, there is a need to develop theoretical tools for analyzing the convergence and stability of RL-augmented quantum optimization loops. This includes:
\begin{itemize}
    \item Establishing bounds on the regret of RL policy in non-stationary environments induced by quantum noise.
    \item Characterizing the stability of Lagrange multiplier updates under stochastic QUBO evaluations.
    \item Investigating whether RL-predicted penalty initialization leads to provably faster convergence in ALM.
\end{itemize}

\bigskip
In summary, this work opens a new research frontier at the intersection of reinforcement learning, quantum computing, and constrained combinatorial optimization. Progress along the above directions can help transform quantum optimization from a theoretical curiosity into a practical tool for next-generation logistics and supply chain systems.

\bibliography{reference}

\end{document}